\documentclass[aps,amsmath]{revtex4}
\usepackage[capitalize]{cleveref}

\usepackage{etoolbox} 
\usepackage{lipsum} 

\makeatletter
\appto{\appendix}{%
  \@ifstar{\def\theequation@prefix{A.}}%
          {}%
}
\makeatother

\usepackage{graphicx}
\usepackage{tikz}
\usetikzlibrary{arrows,shapes,positioning}
\usetikzlibrary{decorations.markings}
\tikzstyle arrowstyle=[scale=1]
\tikzstyle directed=[postaction={decorate,decoration={markings,
    mark=at position .70 with {\arrow[scale=1.5,arrowstyle]{stealth}}}}]
\tikzstyle reverse directed=[postaction={decorate,decoration={markings,
    mark=at position .30 with {\arrowreversed[scale=1.5,arrowstyle]{stealth};}}}]
\newcommand{\be}{\begin{equation}}
\newcommand{\ee}{\end{equation}}
\newcommand{\bea}{\begin{eqnarray}}
\newcommand{\eea}{\end{eqnarray}}

\begin{document}

\title{Two-Agent Games on Graphs}
\author{Tongu\c{c} Rador}
\email[]{tonguc.rador@boun.edu.tr}
\affiliation{Bo\~{g}azi\c{c}i University Department of Physics, 34342 Bebek, \.{
I}stanbul, Turkey\\}

\date{\today}
\begin{abstract}
We study the dynamics of evolution of points of agents placed in the vertices of a graph within the rules
of two-agent units of competition where an edge is randomly chosen and the agent with higher points 
gets a new point with a probability $p$. The model
is closely connected to generalized vertex models and anti-ferromagnetic Potts models at zero temperature.
After studying the most general properties for generic graphs we confine the study to discrete d-dimensional
tori. We mainly focus on the ring and torus graphs.
\end{abstract}
\maketitle

\section{Introduction}

Mathematical modelling is becoming more and more important for problems not only in natural but for  economical and social sciences in general \cite{rev1,rev2}. In this work we study the dynamics of a  model which rests on two-agent competitions first introduced in \cite{red1}; the model has attracted attention and has found various implementations \cite{rede1}-\cite{rede10}. The model was later generalized to 3-agent games \cite{3pap,rad1} and it also created a good
laboratory for studying effects of mergers \cite{rad2}. The 4-agent version of the model was extensively studied in \cite{der1}.

The works on the mentioned model focused exclusively on the implementation of the game on the complete graph which would constitute a mean field approximation for results on generic graphs. In this work we apply the model to general graphs and study various of its properties. What one can say with generality is that the model is a generic vertex model on graphs, and hence is also related to a spin 1/2 system where spins are living on the edges of the graph. One can also say with generality that in the competitive limit the model  is mapped to a constrained graph colouring problem. In that sense it is also kin to anti-ferromagnetic Potts models at zero temperature with constraints on the configurations. Furthermore because the system is a generic vertex model
its dynamics is also described within the mathematical field of hyperplane arrangements; the hyperplanes being defined with
neighbouring vertices.

In II we introduce the model and study its general properties. In III we  study the hydrodynamical approximation and discuss its properties. In IV apply our findings to some finite graphs, including the complete graph. Section V is reserved for an in depth study of the model on the ring graph, where we show that the configurations are 3-colourings of the ring such that there are checkerboard islands of two colours separated by one colour. The size distribution of these islands and the correlation function of the colourings are also studies. We have also implemented a chemical potential approach to study the statistics of the model and show that it is quite accurate even for the details of the model. In section VI we study the model on the two dimensional discrete torus and show that the sizes and perimeters of the mentioned checkerboard islands follow a scaling law. In section VII we digress  on the behaviour of the model for d-dimensional tori. 

\section{The Model and its General Properties}
Let us assume we have $N$ agents occupying vertices on a connected graph with $E$ edges and with no bubbles  so that an agent is not connected to itself by an edge. We label their position with an integer $i=1,2,\cdots,N$ and endow them with integer points $X^{i}$. Let us now randomly pick an edge on an equal likelihood basis. This unambiguously defines two agents. Say, their points are $L$ and $S$ respectively and obey the ordering $L\geq S$. We give a single point to one and only one of them according to the following rules

\bea
(L,S)& \to & (L+1,S)\;\; {\rm with\;\; probability}\;\;p,\nonumber\\
(L,S)& \to & (L,S+1)\;\; {\rm with\;\; probability}\;\;q=1-p,\nonumber
\eea

\noindent whenever $L>S$. If, on the other hand $L=S$, we shall still grant the point, but now on an equal likelihood basis;

\bea
(L,S)& \to & (L+1,S)\;\; {\rm with\;\; probability}\;\;1/2,\nonumber\\
(L,S)& \to & (L,S+1)\;\; {\rm with\;\; probability}\;\;1/2.\nonumber
\eea

Since we have $N$ agents and they have integer points, the state of the system can be described with a vector $\vec{X}=X^{i}\hat{e}_{i}$ in an $N$ dimensional cubic lattice vector space with an orthonormal basis $\hat{e}_{i}$. Note that since we always give a single point after a single unit
of competition, after $k$ such steps the state of the system obeys,

\be
\sum_{i=1}^{N}X^{i}(k)=X_{o}+k,
\ee
where $X_{o}=\sum_{i}{X_{o}}^{i}$ is the sum of the points for the initial state.

The problem at hand is to find $P(\vec{x},k)$ which is the probability at step $k$ that the random variable $\vec{X}$ has value $\vec{x}$ given that at $k=0$ one has the distribution $P(\vec{x},0)$. If the state vector at step $k+1$ is $\vec{X}$, it must have been at a neighbouring vector at step $k$ according to the rules above. We can therefore write down the following master equation for the probability distribution

\be{\label{eq:mas}}
{\mathcal{P}}(\vec{x},k+1)=\sum_{i=1}^{N}{\mathcal{P}}(\vec{x}-\hat{e}_{i},k)v_{i}(\vec{x}-\hat{e}_{i})\;.
\ee

\noindent Where

\be
v_{i}(\vec{x})=\frac{1}{E}\sum_{j\in\mathcal{N}_{i}}\;\left[p\;\theta\left(\vec{x}\cdot(\hat{e}_{i}-\hat{e}_{j})\right)+q\;\theta\left(\vec{x}
\cdot(\hat{e}_{j}-\hat{e}_{i})\right)\right]
\ee

\noindent is, determined by  the microscopic rules, the probability that agent $i$ wins under the condition that one has $\vec{X}=\vec{x}$. In the above $\mathcal{N}_{i}$ is the set of agents neighbouring agent $i$ excluding agent $i$ itself. Also we define ${\rm card}(\mathcal{N}_{i})\equiv d_{i}$, the degree of the vertex associated with $i$'th agent. We use $\theta(0)=1/2$ to implement the tie breaker and also one has 
$\theta(u)+\theta(-u)=1$. Since $p+q=1$, using the handshaking theorem which states that $\sum_{i} d_{i}=2E$, one gets

\be
\sum_{i=1}^{N}v_{i}(\vec{x})=1.
\ee

One can thus recast $v_{i}$ as follows,

\be
v_{i}(\vec{x})=q\frac{d_{i}}{E}+\frac{p-q}{E}\sum_{j\in\mathcal{N}_{i}}\theta(x_{i}-x_{j}).
\ee

We note that for $p=q=1/2$ one has $v_{i}=qd_{i}/E$ and the impact of step functions which depend on the ordering 
of points is absent and $\vec{v}=\sum_{i}\hat{e}_{i}v_{i}(\vec{x})$ is the same constant vector everywhere in the configuration space. 

There are two essential symmetries of the object $\vec{v}$ which we would like to call the {\em velocity field}. These are expressed as

\begin{subequations}
\begin{eqnarray}
\vec{v}(\alpha \vec{x})&=&\vec{v}(\vec{x}),\\
\vec{v}(\vec{x}+\alpha\hat{\mathcal{E}})&=&\vec{v}(\vec{x}).
\end{eqnarray}
\end{subequations}

Here

\begin{equation}
\hat{\mathcal{E}}=\frac{1}{\sqrt{N}}\sum_{i=1}^{N}\hat{e}_{i}
\end{equation}

\noindent is a unit vector along the body diagonal in the $N$ dimensional configuration space.

The velocity field has other properties related to the fact that its dependence on $\vec{x}$ is only via step functions. 
Let us consider the hyperplane $x_{i}=x_{j}$. The microscopic rules can be used to show that 
\be\label{eq:ave1}
\vec{v}(\vec{x}\vert x_{i}=x_{j})=\frac{1}{2}\left[\vec{v}(\vec{x}\vert x_{i}>x_{j})+\vec{v}(\vec{x}\vert x_{i}<x_{j})\right].
\ee

One can show that this averaging property holds also for the intersections of the hyperplanes and
the intersections of the intersections and so on. Of course for a given intersection one will have to average over all hyperplanes, intersections of the hyperplanes and so on that are connected to the intersection.

Let us again consider two agents $i$ and $j$ that are linked. Since a hyperplane will be of dimension $N-1$, the unique normal to it is along the vector $\hat{e}_{i}-\hat{e}_{j}$; that is the perpendicular component is $v_{i}-v_{j}$. One can thus show that the velocity field obeys 
the following boundary conditions 
\begin{subequations}
\begin{eqnarray}
{v_{\parallel}}^{+}-{v_{\parallel}}^{-}&=& 0,\\
v_{\perp}^{+}-v_{\perp}^{-}&=&2\frac{(p-q)}{E},
\end{eqnarray}\end{subequations}
where the superscripts $\pm$ refer to $x_{i}>x_{j}$ and $x_{i}<x_{j}$ respectively.

Various results are implied from the discontinuity of $v_{\perp}$. For $p>q$
we realize two possibilities: Either in both sides the velocity field is pointing away from the hyperplane or in one
side it is pointing towards the hyperplane and on the other pointing away such that the one that points away has larger magnitude, which means that for both cases the hyperplane can only be a transient location. For $p<q$ things are a bit  more involved. We still have the case where in one part velocity field is pointing
away and the other pointing towards such that the one pointing towards has larger magnitude; the opposite situation on the first case for $p>q$. But we also have the possibility
that in both regions the velocities are towards the hyperplane. So for $p<q$ we have the possibility that the hyperplane may be recurrent. However, we have to remind the reader that a hyperplane is never absolutely stable in view of the fact that we have a tie breaker rule in the microscopics of the model which ensures that one can not indefinitely stay on the hyperplane. But the velocity field on the regions
between the hyperplanes might enforce an eventual return to it.   So for $q>p$ we may have agents condensed around the hyperplanes, coming in and out from both regions. Pertinent  to this, is the case where  $v_{\perp}^{-}=-v_{\perp}^{+}$ which would mean that on the hyperplane $x_{i}=x_{j}$ the perpendicular component of the velocity field vanishes.

Another possibility is to have $v_{\perp}$ vanishing. It is clear, due to the discontinuity in $v_{\perp}$, that $v_{i}=v_{j}$ can  occur only in one region 
separated by the $x_{i}=x_{j}$ hyperplane. The condition for $v_{i}=v_{j}$ is the requirement that
\be
q(d_{i}-d_{j})+(p-q)(l_{i}-l_{j})+(p-q)(\epsilon_{i}-\epsilon_{j})/2=0,
\ee
where $l_{i}$ is the number of neighbours of $i$'th agent that has point strictly less than $x_{i}$. Also $\epsilon_{i}$ is the number of neighbours of agent $i$ with which agent $i$ has the same point. Surely in a region where there is an unambiguous ordering of points all $\epsilon_{i}$ are equal to zero. 
This condition generally depends on $q$ if the graph is rather generic. But if we have a {\em regular} graph where 
$d_{i}=d_{j}$ the condition does not depend on $q$ anymore
\;\footnote{Except for the case $q=1/2$ when it is identically satisfied.}. Now if $v_{i}=v_{j}$ the velocity field in this region will be entirely parallel to the hyperplane defined by $x_{i}=x_{j}$.
However due to the discontinuity in $v_{\perp}$, on the other side of the hyperplane the velocity field will definitely point away from the hyperplane for $p>q$ and toward the hyperplane for $p<q$.

One last bit of information about the velocity field is that since by construction all $v_{i}\geq 0$, nowhere the velocity field points towards any $x_{i}=0$ hyperplane. This is also evident from the fact that within the microscopic rules there
is no lowering of points; at any step every agent either stays at the same point or increases its point.

\subsection{A buckets-pipes analogy for the model}

\begin{figure}[t]
\includegraphics[scale=0.3]{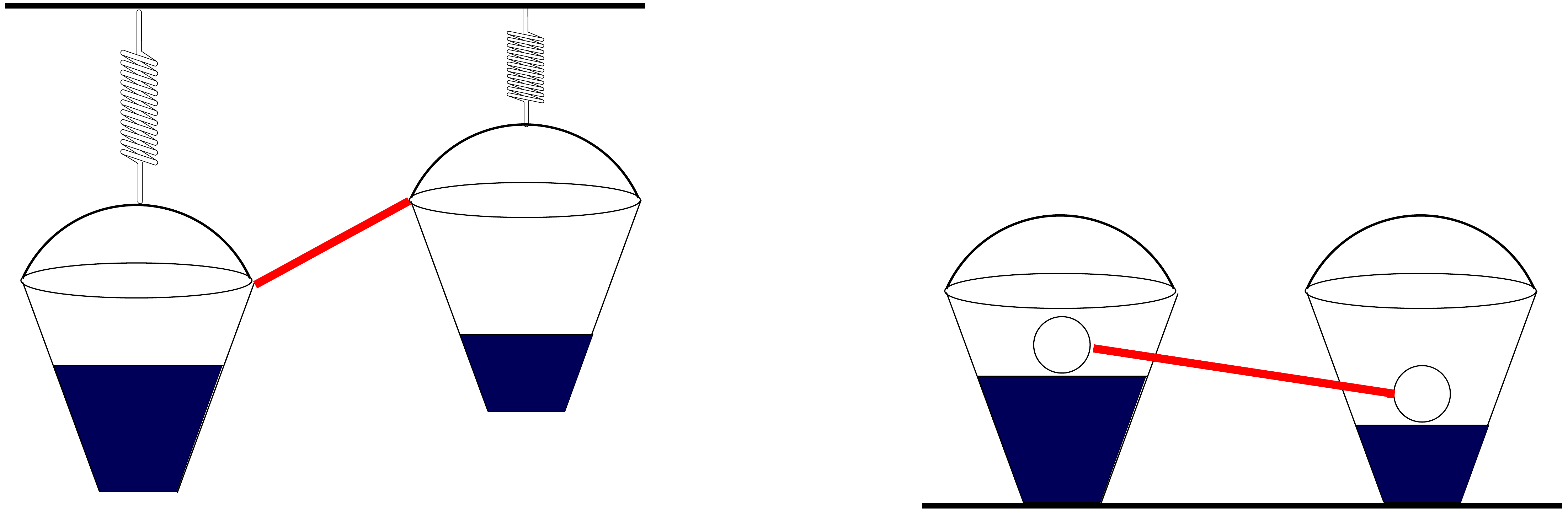}
\caption{The interaction of two agents connected with a link in the buckets-pipes analogy to our model for the two extreme cases $p=1$ and $p=0$. The diagram on the
left represents $p=1$ case where having more amount of water is favoured. The diagram on the right is the
other extreme where having more water is punished. The red lines in the figure represent the pipe where water
is dripped and the blobs are buoys that float on the surface of the water.}
\label{fig:buckets}
\end{figure}

We can form an analogy of the model for $p=0$ or $p=1$ using a network of buckets of water connected with pipes presented in  Fig.\ref{fig:buckets}. 
The buckets represent the agents,
the pipes represent the links and the water level at each pipe represent the points collected. 
When we pick a link we shall drop a defined unit of water on the pipe and
let it flow to one of the buckets. If the water level is the same for two neighbouring buckets we force the
water flow randomly to only one to comply with our rules \footnote{One could, in this case, split the water evenly to
both buckets. However giving points to ties do not change the general structure of the model.}. 

For $p=1$ when we pick a link the agent that has more water in the bucket is to surely receive
the unit volume of water. So we can visualize the system as follows. Assume all buckets are connected to the ceiling with linear elastic bands and the buckets are connected to each other by pipes in a way described by the graph. Let us also assume that the pipes are connected to the top of the buckets. Now
let us start with a given water level distribution for the buckets and drip water on one of the pipes chosen at random and assume that the water will flow instantaneously or that we wait until the system balances via the presence of the elastic bands until we pick another pipe. Under the action of gravity the water will flow to the bucket that has more water and this establishes the analogy. The case $p=1$ however seems to have a more natural connection to competitive systems of
interacting economical entities. The dropping of water will simply stand for taking a share from the overall production.

For $p=0$ having a larger point then one's neighbours must be punished. The way to implement this
behaviour is to put the buckets on the floor and connect the neighbouring ones again with pipes. However this
time do not let the pipes to have a fixed position with respect to the buckets but rest them on the
water surfaces of the buckets with buoys which enforces the water flow under the action of gravity to the bucket with
less water establishing the analogy.

This analogous model is very useful in assessing the qualitative aspects of our system. For instance, if there are cycles in  the graph it gives an understanding that some agents {\em might} still have different water accumulation speeds
even for $p=0$; since there can be a water flow trapped in a portion of the graph. If on the other
hand we are talking of a tree graph the flow seems to always disperse and at $p=0$ all agents travel with
the same rate of water level increase. These statements will be justified later in the manuscript. 
Furthermore the case for $p=0$ may find applications for water systems that are connected with
each other, such as underground caves; the dropping of water in this case simply being representing rainfall.
If on the other hand we still want to make an analogy to interacting entities in an economical system, the case $p=0$ resembles the case for a "rich don't get richer" model. Yet as we have mentioned the number of connections an agent has is also relevant, particularly in an economical system. An inequality in the number of neighbours  generally tends to upset the egalitarian aspect for $p=0$ if there are cycles in the graph.

\section{The hydrodynamical limit}

In the large time and hence large points limit one might contemplate a continuum approach akin to representing the model with a hydrodynamical fluid. In this approximation the discrete equation for $\mathcal{P}$ becomes 

\be\label{eq:cont1}
\frac{\partial {\mathcal{P}}}{\partial t}=-\vec{\nabla}\cdot\left(\vec{v}{\mathcal{P}}\right)+\frac{1}{2}\sum_{i=1}^{N}{\nabla_{i}}^{2}\left(v_{i}\mathcal{P}\right)+\cdots.
\ee

The first term in the equation above describes a convective flow and the second is responsible for diffusion. Whenever 
{\em justifiable} one might ignore diffusion next to convection at very late times. Later in the manuscript, we shall elaborate on what we mean by justifiable 
when we discuss how the diffusion terms, and even sometimes all the terms in the expansion, may become important. For now we simply state that whenever the velocity field in a region points away from all the relevant equal points hyperplanes that bounds it the dominant term is the convective term.
Focusing thus on the first term we end up with

\be\label{eq:contlow}
\frac{\partial {\mathcal{P}}}{\partial t}=-\vec{v}\cdot\vec{\nabla}\mathcal{P}-\left(\vec{\nabla}\cdot\vec{v}\right)\mathcal{P}.
\ee

This is a linear hyperbolic equation and the general method of approach to solve it is that of characteristics. To that end one needs to solve paths in the configuration space such that along them the equation is satisfied. If those paths are described by a parameter $s$ we will have to solve the following set of equations

\begin{subequations}
\begin{eqnarray}
\frac{d\vec{x}(s)}{ds}&=&\vec{v}(\vec{x}(s)),\\
\frac{dt}{ds}&=&1,\\
\frac{d\mathcal{P}(\vec{x}(s),t(s))}{ds}&=&-(\vec{\nabla}\cdot\vec{v})\mathcal{P}.
\end{eqnarray}
\end{subequations}

The last equation in the list above is  the {\em material derivative} of the density function in the fluid dynamics context and will be responsible in assessing the type of the
flow. But since in the regions separated by relevant hyperplanes $\vec{v}$ is  a constant vector field, its divergence is non-zero only along hyperplanes where neighbouring agents have equal points. 

So having focused on the first non-trivial term in our master equation, what one can say is that, in the bulk regions defined between these hyperplanes the dynamics of our model is essentially that of an {\em incompressible flow}.

On the other hand, the divergence of $\vec{v}$ is given as

\begin{equation}
\vec{\nabla}\cdot\vec{v}=\frac{p-q}{E}\sum_{i}\sum_{j\in\mathcal{N}_{i}}\delta(X_{i}-X_{j}),
\end{equation}

\noindent and thus its signature is solely determined by the signature of $p-q=2p-1$. For $p>1/2$ this coefficient is positive and the density on the hyperplanes is diluted; the flow generally points away from them. If on the other hand $p<1/2$
the density on the hyperplanes is condensed; the flow generally tends toward
them where the essentials depend heavily on the particulars of the graph.

So on quite general grounds one can say that for $p>1/2$ nearest neighbours have points that diverge from each other in the infinite time limit and thus the model is related to graph colouring where the colors are the asymptotic speeds.  However what one can  say with confidence for $p<1/2$ case is only that the nearest neighbours {\em might} have converging points and hence same speeds since this might be locally frustrated  by the distribution of the number of neighbours $d_{i}$ on the graph. To make this point clear and elaborate on a point mentioned before, let us ask some questions: For $p=0$ does every agent for a given graph travel at the same   asymptotic speed? If yes what was the critical value of $p$ when this is reached? If not, at $p=0$, what is the number of possible different speeds for a given graph? We shall have partial answers to these later in the manuscript.

\subsection{Expectation values}

Let us define the following 

\begin{equation}
\langle F(\vec{x})\rangle(k)\equiv\sum_{\vec{x}}\;\mathcal{P}(\vec{x},k)F(\vec{x}).
\end{equation}

Using the master equation it is  easy to show, after a shift in the summation variables, that

\begin{equation}
\langle F(\vec{x})\rangle(k+1)=\sum_{i}\langle v_{i}(\vec{x})F(\vec{x}+\hat{e}_{i})\rangle(k).
\end{equation}

The expectation values of $\vec{x}$ and of $\vec{v}(
\vec{x})$ are of central importance. We have

\begin{subequations}
\begin{eqnarray}
\langle\vec{x}\rangle(k+1)-\langle\vec{x}\rangle(k)&=&\langle\vec{v}(\vec{x})\rangle(k),\\
\langle\vec{v}(\vec{x})\rangle(k+1)&=&\sum_{i}\langle\vec{v}(\vec{x}+\hat{e}_{i})v_{i}(\vec{x})\rangle(k).
\end{eqnarray}
\end{subequations}

Note that the equation for the expectation value of the state vector directly follows from the equation obeyed by the random variable $\vec{X}$ which is 

\begin{equation}
\vec{X}(k+1)=\vec{X}(k)+\vec{v}(\vec{X}(k)).
\end{equation}

On the other hand keeping only the convective flow term  in a continuum approach, and using $t$ instead of $k$ we shall get
\begin{equation}
\frac{d\langle F\rangle}{dt}=\langle\vec{v}\cdot\vec{\nabla}F\rangle.
\end{equation}

The expectation values of $\vec{x}$ and $\vec{v}$ in this approximation are,

\begin{subequations}
\begin{eqnarray}
\frac{d\langle\vec{x}\rangle}{dt}&=&\langle\vec{v}\rangle,\\
\frac{d\langle\vec{v}\rangle}{dt}&=&\langle\vec{v}\cdot\vec{\nabla}\vec{v}\rangle.
\end{eqnarray}
\end{subequations}

Note that the expectation value of $\vec{x}$ is already saturated by the convective approximation.
Also, as in fluid dynamics, the quantity $\vec{v}\cdot\vec{\nabla}\vec{v}\equiv\vec{a}$ is the acceleration field. Its components are given as

\begin{equation}
a_{i}=\frac{p-q}{E}\sum_{j\in\mathcal{N}_{i}}(v_{i}-v_{j})\delta(X_{i}-X_{j}).
\end{equation}

Evidently the acceleration is only non-zero when some neighbouring agents have the same points and having different speeds. Thus in the bulk regions, where no neighbouring agents have the same points, the acceleration vanishes and the expectation value of the velocity field converges to a constant value. Furthermore one can still have a vanishing acceleration on a neighbourhood hyperplane $x_{i}=x_{j}$ if 
$v_{i}=v_{j}$, as mentioned before.

Therefore

\begin{equation}
\lim_{t\to\infty}\frac{d\langle\vec{v}\rangle}{dt}\to 0,
\end{equation}
generally in the bulk regions and also on an hyperplane if the perpendicular component vanishes. 

This in turn means that

\begin{equation}
\lim_{t\to\infty}\langle\vec{X}\rangle\to\langle\vec{v}\rangle\;t+\langle\vec{X}\rangle(t=0).
\end{equation}

\noindent As an example if one picks $\mathcal{P}(\vec{X},0)=\delta^{N}(\vec{X})$ one shall get $\langle\vec{X}\rangle(t=0)=0$. In fact if we assume that the initial distribution has {\em finite moments} we can generally say that

\begin{equation}
\lim_{t\to\infty}\langle\vec{X}\rangle\to\langle\vec{v}\rangle\;t.
\end{equation}

To reiterate let us know focus on $\langle X_{a} X_{b}\rangle$. It is easy to show that

\begin{equation}
\frac{d\langle X_{a}X_{b}\rangle}{dt}=\langle v_{a}X_{b}\rangle +\langle v_{b}X_{a}\rangle.
\end{equation}

Taking a further derivative and focusing on the bulk regions we can arrive at

\begin{equation}
\frac{d^{2}\langle X_{a}X_{b}\rangle}{dt^{2}}=2\langle v_{a}v_{b}\rangle.
\end{equation}

However as we have shown that the acceleration field vanishes in the bulk regions the quantity $\langle v_{a}v_{b}\rangle$ approaches a constant in the infinite time limit, and we thus expect the following if
the initial distribution of points has finite moments

\begin{equation}
\lim_{t\to\infty}\langle X_{a}X_{b}\rangle\to t^{2}\langle v_{a}v_{b}\rangle.
\end{equation}

In fact by assuming finite moments for the initial distribution the following general relation holds in the bulk regions of the configuration space for the asymptotic dynamics

\be{\label{eq:momen22}}
\lim_{t\to\infty}\langle \prod_{k=1}^{K} X_{i_{k}}^{p_{k}} \rangle (t)\to\langle \prod_{k=1}^{K} v_{i_{k}}^{p_{k}} \rangle t^{S},
\ee
 
\noindent where $S=\sum_{k}p_{k}$.
\subsection{The scaling symmetry of $\vec{v}(\vec{x})$ and the $\mathcal{P}(\vec{x}/t)$ ansatz}

These facts above are closely related to the scaling symmetry of the velocity field. If we remember that
$\vec{v}(\vec{x}/t)=\vec{v}(\vec{x})$ we see that the equation governing $\mathcal{P}$ will allow
an asymptotic  solution of the form $\mathcal{P}(\vec{x}/{t})$ provided the initial condition does not spoil it. Such types of initial conditions are for instance starting all agents from the same point, or say sampling initial conditions from inside an $N$-dimensional cube or sphere in the configuration space on an equal likelihood basis.

Defining $\vec{z}\equiv\vec{x}/t$, the equation in the bulk regions will have the form

\be{\label{eq:eqforz}}
\left[-\vec{z}+\vec{v}(\vec{z})\right]\cdot\nabla_{\vec{z}}\mathcal{P}(\vec{z})=0.
\ee

This can also be approached via the method of characteristics. The system one will have to solve is the
following

\begin{subequations}
\begin{eqnarray}
\frac{dz_{i}(s)}{ds}&=&-z_{i}+v_{i}(\vec{z}),\\
\frac{d\mathcal{P}}{ds}&=&0.
\end{eqnarray}
\end{subequations}

Since in the asymptotic regime we have found that the random variable $\vec{X}/t$ approaches a constant, which is $\vec{v}$, we arrive at the defining relation for an asymptotic solution

\begin{equation}{\label{eq:voila}}
\boxed{\vec{z}=\vec{v}({\vec{z}})}
\end{equation}
along with
\begin{equation}
\boxed{\sum_{i}z_{i}=1}
\end{equation}

\noindent and the probability that a configuration satisfying the condition above is a constant\footnote{A humourous person
would write Eq.\ref{eq:voila} as $\vec{v}=\vec{v}(\vec{v})$.}. Also we note that since the velocity field depends only on step functions which take only two values, the configurations defined by Eq.\ref{eq:voila} are those  of a generalized vertex model on the graph.

So what we have to do has two aspects: One has to assess the list of possible configurations that satisfy Eq.\ref{eq:voila} and compute probabilities of occurrence for these configurations. The list of configurations  depends only on the graph properties via the velocity field and generally on $q$ but the asymptotic occurrence probabilities of these will in general depend on the initial conditions. Thus we expect broken ergodicity on general grounds.

Let us now remember that the exact behaviour of the expectation value of $\vec{x}$ is the same as in the
lowest order approximation in the hydrodynamical approximation. Therefore Eq.\ref{eq:voila} will also
be valid for $p<q$ and this means we can find all the possible speeds and hence possible configurations
of the system for all $p$. The generic recipie is that we look for all acyclic orientations of the graph that will yield a colouring for $p>q$; some will provide solutions and some will not. For $p<q$ however we are allowed to let neighbouring
agents have the same speeds and one has to also look for those possibilities; that is say {\em partial orientations}. Again, some of these may provide solutions and some may not. We shall elaborate more on the generalalities of Eq.\ref{eq:voila} later in the manuscript.

\subsection{Universality classes $p\leq 1/2$ and $p>1/2$ for regular graphs}

For regular graphs there is enormous simplification of Eq.\ref{eq:voila}. Since on a regular graph the number of neighbours are the same, say $d$, we have

\be
z_{i}=\frac{qd}{E}+\frac{p-q}{E}\sum_{j\in\mathcal{N}_{i}}\theta(z_{i}-z_{j}).
\ee

Due to the nature of the step function the following transformation
\be
\zeta_{i}\equiv (z_{i}-\frac{qd}{E})\frac{E}{p-q}
\ee
will bring the equation to the form
\be
\zeta_{i}=\sum_{j\in\mathcal{N}_{i}}\theta(\epsilon(\zeta_{i}-\zeta_{j})),
\ee
where $\epsilon=\rm{Sign}(p-q)$. This hints at  two universality classes of different values of $p$ separated only by the signature of $p-q$.

To reiterate on the property just mentioned, let us go back to the original master equation. In the continuum limit keeping only the convective flow terms one can show that it can be written as

\be
\frac{\partial \mathcal{P}}{\partial t}=-q\vec{D}\cdot\vec{\nabla}\mathcal{P}-(p-q)\vec{\nabla}\cdot\left(\vec{\omega}\mathcal{P}\right),
\ee
 where $\vec{D}=\sum\hat{e}_{i}d_{i}/E$ is the vertex degree vector and $\omega_{i}=\sum_{j\in\mathcal{N}_{i}}\theta\left(\vec{x}\cdot(\hat{e}_{i}-\hat{e}_{j})\right)/E$. Surely $\vec{\omega}$ is the same vector as $\vec{v}$ for $p=1$. 

For a regular graph one has $D_{i}=d=2E/N$. Let us also remember that the velocity field enjoys a scaling symmetry and a shift symmetry if the shift is along $\hat{\mathcal{E}}$. Thus under the following transformation 

\[
\vec{X}\to\frac{\vec{X}-qt\vec{D}}{p-q},
\]
the velocity field is unchanged. Consequently the equation can be transformed to 

\be
\frac{\partial \mathcal{P}}{\partial t}=-\vec{\nabla}\cdot\left(\vec{\omega}\mathcal{P}\right)
\ee
Which is the same equation as our original starting point had we picked $p=1$. Thus the argument for the mentioned universality classes does not depend on the $\mathcal{P}(\vec{x}/t)$ ansatz.

To summarize, for regular graphs in the $p\leq 1/2$ regime all agents have the same asymptotic speed of $1/N$ and for $p>1/2$ the problem is statistically equivalent to the case $p=1$. 

Of course in an actual simulation of the system if $p\neq 1$ but still $p>1/2$ the convergence of the system to the statistics described by $p=1$ should take longer, a fact we have substantiated in simulations. The reason is that the characteristics diverge from each other in a slower way if $p\neq 1$. Thus this universality is welcome and to be exploited to make simulations run  faster since for $p=1$ one does not have to pick a random number to assess which of the two neighbours that were picked will win the game.

Another property of  regular graphs is that for all of them one can consider a thermodynamic limit  where the number of agents is taken to infinity while keeping the properties of the graph intact. 

\subsection{Connection to hyperplane arrangements and graph colourings}

The step functions in the velocity field  divide the configuration space into regions where the velocity field is constant but have strengths and directions that depend on the region. This makes an immediate connection to a branch in combinatorial geometry called \textit{hyperplane arrangements}. In particular since the arrangements of these hyperplanes depend on a given graph they fall into the category of \textit{graph arrangements}. For definitions and theorems on this particular topic which we shall cite below we refer the reader to \cite{orlik} and  \cite{stanley}.

The  arrangements in our model are \textit{central} ones, meaning that  all hyperplanes include the origin. Because of this it is easy to show  that they are all unbounded using Zaslavsky's theorem: They bound each other but they all have an open end towards infinity in  our point configuration space. This is to be expected from the general behaviour of our  model since the points of  agents increase indefinitely and no agent loose points.

The theory of graphic arrangements have close relations to graph colouring problem in that 
the defining polynomial of the arrangement is given by the chromatic polynomial of the graph. The chromatic polynomial $\chi_{G}(q)$ of a graph counts the number of ways a graph $G$ can be coloured with $q$ colours. In our context, colour means the asymptotic point gain speed of the agent in view of Eq.\ref{eq:voila}, for $p>1/2$.

For a graph $G$ with $N$ vertices, the number of \textit{chambers} which is the technical term used for the bulk regions defined by the velocity field where no neighbouring agent has the same points, is given by $(-1)^{N}\chi_{G}(-1)$. Also this number is equal to the number of acyclic orientations of $G$. An orientation of a graph is obtained by assigning directions on the links of the graph that joins the vertices and the total number of such orientations is simply $2^{E}$. An acyclic orientation is one that does not create oriented cycles in the diagram and thus, in our context, defines an instance of a local absolute ordering of points of the agents residing on the vertices \footnote{If the orientation can be cyclic it can not be associated with an ordering of points for those agents residing on nodes of the cycle in question, they will have equal points in a cyclic orientation.}.

Using this language the meaning of Eq.\ref{eq:voila} is clearer: For $p>1/2$ it's solutions are colourings that are defined by an
acyclic orientation such that the speed, that is the colour, is equal to the number of outgoing arrows from a vertex. Note that not all acyclic orientations will therefor provide colourings that will resolve Eq.\ref{eq:voila} and conversely not all colourings
obtained using the maximum number of colours defined with $d_{\rm max}$ will provide a meaningful acyclic orientation. The constraint equation Eq.\ref{eq:voila} is non-trivial in this respect.

\subsection{Roles of the non-convective terms}

The non-convective terms dominate the evolution for $p<1/2$ since here the points are generally drawn near the hyperplanes via the convective flow. The dynamics in this case is a complicated pattern of agents hitting a hyperplane then diffuse out only to return back in short time via the convective flow. The discussion we have below is therefore confined to  the case $p>1/2$ where the convective terms tend to push the state away from the hyperplanes. However, as we have already mentioned and as we shall discuss with explicit examples the solution to the pure convective result in Eq.\ref{eq:voila} can be used to assess the configurations of asymptotic speeds even for $p<1/2$. 

There are two major roles of the non-convectice terms. First they dominate the evolution of the system for early times: In a simulation where the actual configuration space is discrete one possibly spends some time early in the simulation not in the bulk regions but on the hyperplanes dividing these regions and even on their further intersections. This of course depends on the inital conditions: If one picks initial conditions from a large cube chances are one will be deep in a well defined
bulk region, but if one starts all agents from the same points early times are dominated purely
by the diffusion. Second, even  when the state
actually falls well in a bulk region the velocity field might be parallel to a hyperplane and thus for $p>1/2$ can not create a flow diverging from that hyperplane\;\footnote{ Remember that we have already shown that when
two connected agents $i$ and $j$ have the same speed, that is $v_{i}=v_{j}$, the velocity field is parallel to the hyperplane $x_{i}=x_{j}$.}.

Now, let us take  two connected agents which we label as $x_{1}$ and $x_{2}$, and let us assume we are in a bulk region with a well defined ordering of points and hence one has a 
constant velocity field throughout. The following co-ordinate transformation 
\bea
u&=& x-y\nonumber,\\
w&=& x+y\nonumber,
\eea
will cast the hydrodynamical form of the master equation to the following 

\be\label{eq:chamberdyna}
\frac{\partial P}{\partial t}=-2D\frac{\partial P}{\partial w}+D\frac{\partial ^{2}P}{\partial w^{2}}+D\frac{\partial^{2}P}{\partial u^{2}}-\sum_{i=3}^{N}v_{i}\partial_{i}P+\frac{1}{2}\sum_{i=3}^{N}v_{i}{\partial_{i}}^{2}P
\ee
where the last two summations contain no $u$ and $w$ dependence inside this chamber.

We realize that the co-ordinate $u$ which is along $\hat{e}_{1}-\hat{e}_{2}$ performs ordinary diffusion. It is well known that the eventual probability for the walker to hit $u=0$, which is $x_{1}=x_{2}$ plane, is 1. Yet the mean time to do so is infinite. Unfortunately. this is a nuissance for actual simulations  which by definition must encompass a finite time. 

We can thus define two types of bulk regions: Ones that we would like to call {\em stable} and the others {\em unstable}. The stable bulk regions are where, for $p>1/2$, the velocity field is pointing away from all hyperplanes bounding that region. The unstable ones are where, again for $p>1/2$, the velocity field is parallel to, possibly more than one, hyperplane. 

To exemplify this behaviour let us assume the following local piece of a graph along with a given local acyclic orientation,

\begin{center}
$\cdots$\begin{tikzpicture}
 
  \node[scale=1,auto=left,style={circle,fill=blue!20}] (n1) at (1,0)  {2};
  \node[scale=1,auto=left,style={circle,fill=blue!20}] (n2) at (2,0) {1};
  \node[scale=1,auto=left,style={circle,fill=blue!20}] (n3) at (3,0)  {1};
  \node[scale=1,auto=left,style={circle,fill=blue!20}] (n4) at (4,0) {0};

 \node[scale=1,auto=left,style={circle,fill=white!20}] (n0) at (0,0) {};
 \node[scale=1,auto=left,style={circle,fill=white!20}] (n5) at (5,0) {};
    \draw[directed] (n1)--(n2);
    \draw[directed] (n2)--(n3);
    \draw[directed] (n3)--(n4);
    \draw[reverse directed] (n0)--(n1);
    \draw[reverse directed] (n4)--(n5);
\end{tikzpicture}$\cdots$
\end{center}

Where the labels are the {\em would be} asymptotic speeds  of the agents which for simplicity we presented for $p=1$ and normalized to the number of outgoing arrows. This example is pertinent to the ring graph which we shall study at length in this manuscript. Clearly this is not a proper colouring of this portion of the graph
and in the bulk region defined with the orientation above the velocity field is parallel to hyperplanes defined with the neighbouring agents which have the same speed.

Now as it stands this configuration must evolve with the diffusing terms since as we have shown the point differences of the agents with the same speed will perform an ordinary random walk. The moment when two neighbouring agents with
the same speed have equal points we shall be on the hyperplane. Thus the configuration may evolve to the following
  
  \begin{center}
$\cdots$\begin{tikzpicture}
 
  \node[scale=1,auto=left,style={circle,fill=blue!20}] (n1) at (1,0)  {2};
  \node[scale=1,auto=left,style={circle,fill=blue!20}] (n2) at (2,0) {0};
  \node[scale=1,auto=left,style={circle,fill=blue!20}] (n3) at (3,0)  {2};
  \node[scale=1,auto=left,style={circle,fill=blue!20}] (n4) at (4,0) {0};

 \node[scale=1,auto=left,style={circle,fill=white!20}] (n0) at (0,0) {};
 \node[scale=1,auto=left,style={circle,fill=white!20}] (n5) at (5,0) {};
    \draw[directed] (n1)--(n2);
    \draw[directed] (n3)--(n2);
    \draw[directed] (n3)--(n4);
    \draw[reverse directed] (n0)--(n1);
    \draw[reverse directed] (n4)--(n5);
\end{tikzpicture}$\cdots$
\end{center}
which constitute a stable configuration, or may fall back to the old unstable bulk region, just shy of the hyperplane.  In the long run this falling back does not change the eventual
probability to cross the hyperplane, albeit with an inifinite mean time to do so.

Of course if the chain with the speed $1$ agents were longer the  transitions are richer and may eventually contain a stable configuration with a number of speed $1$ agents. Consider for instance the following initial configuration

 \begin{center}
$\cdots$\begin{tikzpicture}
 
  \node[scale=1,auto=left,style={circle,fill=blue!20}] (n1) at (1,0)  {2};
  \node[scale=1,auto=left,style={circle,fill=blue!20}] (n2) at (2,0) {1};
  \node[scale=1,auto=left,style={circle,fill=blue!20}] (n3) at (3,0)  {1};
  \node[scale=1,auto=left,style={circle,fill=blue!20}] (n4) at (4,0) {1};
  \node[scale=1,auto=left,style={circle,fill=blue!20}] (n5) at (5,0) {0};

 \node[scale=1,auto=left,style={circle,fill=white!20}] (n0) at (0,0) {};
 \node[scale=1,auto=left,style={circle,fill=white!20}] (n6) at (6,0) {};
    \draw[directed] (n1)--(n2);
    \draw[directed] (n2)--(n3);
    \draw[directed] (n3)--(n4);
    \draw[directed] (n4)--(n5);
    \draw[reverse directed] (n0)--(n1);
    \draw[reverse directed] (n5)--(n6);
\end{tikzpicture}$\cdots$
\end{center}

This will eventually evolve to either the following stable configuration

\begin{center}
$\cdots$\begin{tikzpicture}
 
  \node[scale=1,auto=left,style={circle,fill=blue!20}] (n1) at (1,0)  {2};
  \node[scale=1,auto=left,style={circle,fill=blue!20}] (n2) at (2,0) {0};
  \node[scale=1,auto=left,style={circle,fill=blue!20}] (n3) at (3,0)  {2};
  \node[scale=1,auto=left,style={circle,fill=blue!20}] (n4) at (4,0) {1};
  \node[scale=1,auto=left,style={circle,fill=blue!20}] (n5) at (5,0) {0};

 \node[scale=1,auto=left,style={circle,fill=white!20}] (n0) at (0,0) {};
 \node[scale=1,auto=left,style={circle,fill=white!20}] (n6) at (6,0) {};
    \draw[directed] (n1)--(n2);
    \draw[directed] (n3)--(n2);
    \draw[directed] (n3)--(n4);
    \draw[directed] (n4)--(n5);
    \draw[reverse directed] (n0)--(n1);
    \draw[reverse directed] (n5)--(n6);
\end{tikzpicture}$\cdots$
\end{center}

\noindent or to the following one

\begin{center}
$\cdots$\begin{tikzpicture}
 
  \node[scale=1,auto=left,style={circle,fill=blue!20}] (n1) at (1,0)  {2};
  \node[scale=1,auto=left,style={circle,fill=blue!20}] (n2) at (2,0) {1};
  \node[scale=1,auto=left,style={circle,fill=blue!20}] (n3) at (3,0)  {0};
  \node[scale=1,auto=left,style={circle,fill=blue!20}] (n4) at (4,0) {2};
  \node[scale=1,auto=left,style={circle,fill=blue!20}] (n5) at (5,0) {0};

 \node[scale=1,auto=left,style={circle,fill=white!20}] (n0) at (0,0) {};
 \node[scale=1,auto=left,style={circle,fill=white!20}] (n6) at (6,0) {};
    \draw[directed] (n1)--(n2);
    \draw[directed] (n2)--(n3);
    \draw[directed] (n4)--(n3);
    \draw[directed] (n4)--(n5);
    \draw[reverse directed] (n0)--(n1);
    \draw[reverse directed] (n5)--(n6);
\end{tikzpicture}$\cdots$
\end{center}

Thus the problem here is to find the splitting probabilities of exit to either stable region. This clearly has direct dependence on the initial location inside the unstable region; the very source of broken ergodicity we have mentioned. However if the initial conditions are such that one has equal probability to start from anywhere within the unstable region of type presented above, the splitting probabilities are distributed evenly \footnote{For instance see \cite{redbook1} and references therein.} across the stable regions that surround it.

\subsection{On initial conditions and interpretation of the asymptotics}

The discussion on the role of non-convective terms makes it clear that the general solution, though in principle tractable, has heavy dependence on the initial conditions. If, on top of this, we remember that the convective terms dominate the
long term asypmtotics in a stable chamber for $p>1/2$ we realize that a change in the questions of interest might help.
So let us ask: For $p>1/2$, given that the initial condition is evenly distributed in a particular chamber how does the asymptotic state distributes itself among the stable chambers? The answer is simple. In view of the equality of splitting probabilities if one samples evenly in a bounded region one simply has to count the number of stable chambers which bound the region where the initial state resides. This is a problem in combinatorial geometry that has powerful tools. The solution however will again depend on the particulars of a given graph, and though constituting a simpler problem than the most generic initial condition may still require formidable effort. In view of this one can even generalize slightly: Say we 
randomly sample initial conditions from a cube cornered at the origin or a sphere centered at the origin. When the size of these  grow such that the hyperplane contributions to the volume can be neglected in a discrete lattice, every chamber will sampled in accordance with the volumes they occupy. The calculation of the relative volumes of chambers in a given graph hyperplane arrangement is  fairly straightforward since they are central, but a closed form is not necessarily available for regular graphs of infinite size. The final answer to the question just asked will thus have to include these ratios as well. 

The second type of initial condition that stands out is  one where all agents start at the same point, which can always be chosen to be zero. In this case the early stages of a simulation is  complicated since on a discrete lattice the bulk regions can not be distinguished too close to the origin; there is simply not enough points on the lattice. For this reason many of the agents will be stuck at zero points for $p>1/2$ and thus the ratio of zero speed agents will be larger than the previous type of initial conditions; we are mostly on the hyperplanes defined by the co-ordinate zeros. The asymptotic properties will thus be different.

Once these transition probabilities between chambers is known the problem is mapped to a Markov process and the final distribution
can in priciple be obtained \cite{brown}. Yet obtaining these transition probabilities is a formidable job in general and in this work we give explicity study of them only for a few graphs.

\section{Some examples on finite graphs}

In this subsection we shall study the system on some finite graphs to expose the properties we have so far established.

\subsection{The smallest graph}

\begin{figure}[t]
\includegraphics[scale=1.5]{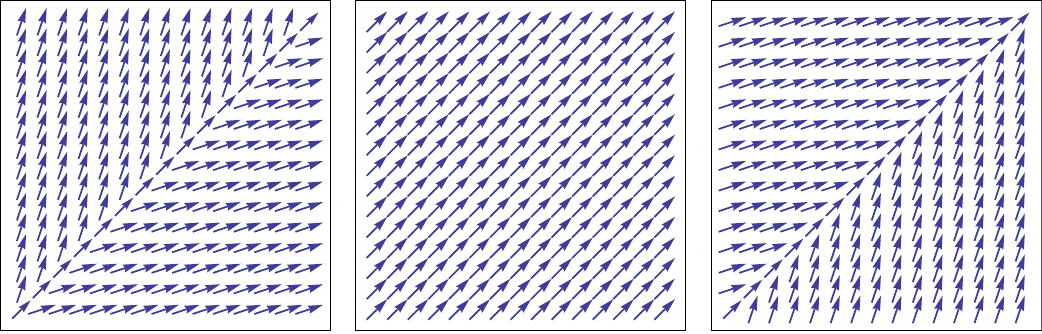}
\caption{The velocity field for the graph with only two vertices. The graph on the left is for $p=3/4$, the one in the middle is for $p=1/2$ and the one on the right is for $p=1/4$. In each case there is a vector along the body diagonal with norm $1/\sqrt{2}$. This line is the only hyperplane present for this graph arrangement.}
{\label{fig1}}
\end{figure}

The smallest non-trivial graph is one that has two vertices and a single edge that connects them; note that this is a regular graph. Let us represent their points on a two dimensional plane. The visualization of the velocity field is presented in Fig.\ref{fig1}.

The simple structure of the graph divides the configuration space into two equivalent parts. We thus infer
that for $p>1/2$ no matter what the initial conditions are one of the agents will have asymptotic speed $p$ and the other $q=1-p$ this property is quite visual from the figure of the vector field. The statistics of the system is equivalent to that of $p=1$ and the probabilities of the asymptotic configurations are equally likely if one starts from the origin since the geometric structure of the bulk regions enjoys a flip symmetry about the body diagonal in the configuration space. For $p\leq 1/2$ the asymptotic speeds of the agents are the same and a shock is formed as the flow converges to the body diagonal as apparent from the figure.

\subsection{Complete graph with $N$ nodes}

In a complete graph every vertex has an edge to every other: That is every agent can play a game with every other and hence this constitutes the basis for the mean field approach to an arbitrary graph. It is also a regular graph. Thus for $p\leq 1/2$ all agents will have the same asymptotic speed and a shock is formed trough the entire graph. The statistical behaviour for $p>1/2$ is identical to that of $p=1$ which has the following velocity field,

\be
v_{i}(\vec{X})=\sum_{j\neq i}\theta(X_{i}-X_{i}),
\ee
where we have normalized so that the sum of its componente is $N$. Since all the agents are linked none will have the same speed and  the application of the consistency relation Eq.\ref{eq:voila} will mean

\be
z_{a}=\sum_{b\neq a}\theta(z_{a}-z_{b}),
\ee

\noindent where we have to deny zero arguments in the step function using our argumentation on the stable and unstable chambers. The only consistent solution to the equation above, within our requirements, is that there is a unique agent with $z=0$, a unique agent with $z=1$, another unique agent with $z=2$, and finally at the end a unique agent with $z=N-1$.

Thus if we label the number of agents with points less than $s$ at time $t$ by $F(s,t)$, we know that it goes to a function of $F(s/t)$ and defining a proper normalization for $z=s/t$ such that the mean velocity of the list of agents is $1/2$ we find that $F(z)=z$ between $0\leq z\leq 1$. This result is well known since the model and some of its extensions on the complete graph are extensively studied. Here the framework we established helps us understand it as {\em the solution to the  colouring problem  on the complete graph with $N$ nodes.} To take a note in the literature: The fact that for $p>1/2$ all agents travel at different speeds for this system was mentioned in the past, albeit not in this form; see Fig.3 in \cite{3pap}.

We see that the constraints imposed by Eq.\ref{eq:voila} are not too spectacular for this graph, since every colouring of a complete graph is a strict ordering. Also, the hyperplane arrangement associated with this graph enjoys the full permutation symmetry and every chamber has equal relative volume when intersected with a $N$-dimensional cube or sphere with the center at the origin. This means asymptotically that every ordering is equally likely to occur if the initial condition does not upset this symmetry. 

\subsection{The N=3 Tree Graph}
We now consider the smallest non-regular graph given as below
\begin{center}
\begin{tikzpicture}
  \node[scale=1,auto=left,style={circle,fill=blue!20}] (n1) at (1,0)  {a};
  \node[scale=1,auto=left,style={circle,fill=blue!20}] (n2) at (2,0) {c};
  \node[scale=1,auto=left,style={circle,fill=blue!20}] (n3) at (3,0)  {b};
    \draw (n1)--(n2);
    \draw (n2)--(n3);
  \end{tikzpicture}
\end{center}
where the labels are those that we shall use to denote  the agents residing
at those vertices. The velocity field is,

\begin{subequations}
\begin{eqnarray}
v_{a}&=&\frac{q}{2}+\left(\frac{p-q}{2}\right)\theta(a-c),\\
v_{b}&=&\frac{q}{2}+\left(\frac{p-q}{2}\right)\theta(b-c),\\
v_{c}&=&q+\left(\frac{p-q}{2}\right)\left[\theta(c-a)+\theta(c-b)\right].
\end{eqnarray}
\end{subequations}

\begin{figure}[t]
\includegraphics[scale=0.5]{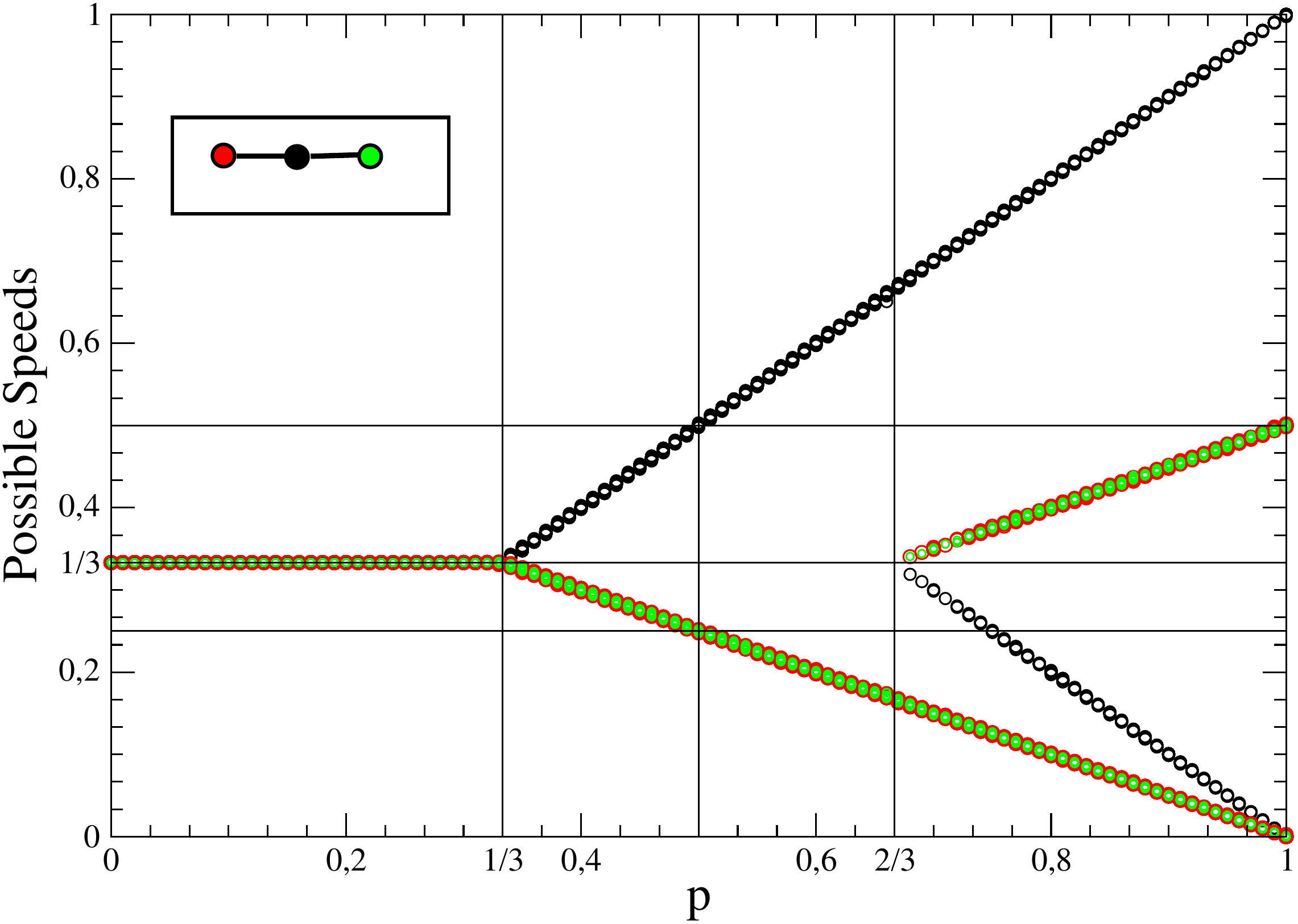}
\caption{The possible asyptotic speeds of agents for the graph shown in the inset. The colour code in the plot matches the color assignment of the agents as shown in the inset.}\label{fig:n3l21}
\end{figure}

We start by looking for all possible asymptotic speeds. To that end we performed simulations  for every value of $p$, the results are presentend in Fig.\ref{fig:n3l21}. 

We shall now show that the structure of possible speeds can be fully understood with the
use of Eq.\ref{eq:voila} even for $p\leq 1/2$. First of all one can show that for this graph no partial orientation solution exists where some neighbouring agents have the same speed; more on this in just a moment. We thus turn to the complete orientations of the graph which are given below

\begin{center}
\begin{tikzpicture}$S_{T}\; :$
  \node[scale=1,auto=left,style={circle,fill=blue!20}] (n1) at (1,0)  {a};
  \node[scale=1,auto=left,style={circle,fill=blue!20}] (n2) at (2,0) {c};
  \node[scale=1,auto=left,style={circle,fill=blue!20}] (n3) at (3,0)  {b};
    \draw[directed] (n2)--(n1);
    \draw[directed] (n2)--(n3);
  \end{tikzpicture}
\end{center}
\begin{center}
\begin{tikzpicture}$S_{B}\; :$
  \node[scale=1,auto=left,style={circle,fill=blue!20}] (n1) at (1,0)  {a};
  \node[scale=1,auto=left,style={circle,fill=blue!20}] (n2) at (2,0) {c};
  \node[scale=1,auto=left,style={circle,fill=blue!20}] (n3) at (3,0)  {b};
    \draw[directed] (n1)--(n2);
    \draw[directed] (n3)--(n2);
  \end{tikzpicture}
\end{center}
\begin{center}
\begin{tikzpicture}$U_{L}\; :$
  \node[scale=1,auto=left,style={circle,fill=blue!20}] (n1) at (1,0)  {a};
  \node[scale=1,auto=left,style={circle,fill=blue!20}] (n2) at (2,0) {c};
  \node[scale=1,auto=left,style={circle,fill=blue!20}] (n3) at (3,0)  {b};
    \draw[directed] (n1)--(n2);
    \draw[directed] (n2)--(n3);
  \end{tikzpicture}
\end{center}
\begin{center}
\begin{tikzpicture}$U_{R}\; :$
  \node[scale=1,auto=left,style={circle,fill=blue!20}] (n1) at (1,0)  {a};
  \node[scale=1,auto=left,style={circle,fill=blue!20}] (n2) at (2,0) {c};
  \node[scale=1,auto=left,style={circle,fill=blue!20}] (n3) at (3,0)  {b};
    \draw[directed] (n2)--(n1);
    \draw[directed] (n3)--(n2);
  \end{tikzpicture}
\end{center}

We can easily show that the acyclic orientations called $U_{L}$ and $U_{R}$ do not provide solutions to Eq.{\ref{eq:voila}, except for the case $p=q=1/2$ where all regions are treated equally anyways, and this justifies calling them unstable chambers. 

The acyclic orientation $S_{B}$ simply demands that one must has $z_{a}=z_{b}=p/2$ and $Z_{c}=q$ and this must be a solution to Eq.{\ref{eq:voila}, meaning
\begin{subequations}
\begin{eqnarray}
z_{a}&=&\frac{p}{2}=v_{a}(\vec{z})=\frac{q}{2}+\left(\frac{p-q}{2}\right)\theta(\frac{p}{2}-q),\\
z_{b}&=&\frac{p}{2}=v_{b}(\vec{z})=\frac{q}{2}+\left(\frac{p-q}{2}\right)\theta(\frac{p}{2}-q),\\
z_{c}&=&q=v_{c}(\vec{z})=q+\left(\frac{p-q}{2}\right)\;2\;\theta(q-\frac{p}{2}).
\end{eqnarray}
\end{subequations}

We clearly see that the solution is valid only for $p>2/3$. If $p$ drops below this value
this acyclic orientation does not provide a solution and hence disappears from the list
of possibly asymptotic speeds as can be clearly seen from the behaviour of the simulations presented in Fig.\ref{fig:n3l21}.

Coming to the case of  $S_{T}$ we realize that the speeds must be $z_{a}=z_{b}=q/2$ and $z_{c}=p$. Since this must solve Eq.{\ref{eq:voila}, we need

\begin{subequations}
\begin{eqnarray}
z_{a}&=&\frac{q}{2}=v_{a}(\vec{z})=\frac{q}{2}+\left(\frac{p-q}{2}\right)\theta(\frac{q}{2}-p),\\
z_{b}&=&\frac{q}{2}=v_{b}(\vec{z})=\frac{q}{2}+\left(\frac{p-q}{2}\right)\theta(\frac{q}{2}-p),\\
z_{c}&=&p=v_{c}(\vec{z})=q+\left(\frac{p-q}{2}\right)\;2\;\theta(p-\frac{q}{2}).
\end{eqnarray}
\end{subequations}

As can be seen the solution is valid for $p>1/3$, as can be seen from Fig.\ref{fig:n3l21}, where there are two distinct speeds for each $p$. 

What happens for $p<1/3$? Well, clearly Eq.{\ref{eq:voila}} can not be satisfied with all equal speeds. However we also have the
constraint that $\sum_{a}z_{a}=1$. Thus below the value where there are no more solutions to Eq.{\ref{eq:voila}} via partial orientatons, the situation is resolved by the fact that all $z_{a}$ are the same and are equal to $1/3$. This situation is in exact analogy to the resolution  of a shock in hyperbolic differential equations via the Rankine-Hugoniot conditions.

We thus realize that Eq.\ref{eq:voila} along with orientations of a given graph is giving enough information about the possible asymptotic speeds of agents. The details of course will depend heavily on the topology of the graph.

These behaviours can also be understood by plotting the velocity field of the graph which we present in Fig.\ref{fig:meem}. To that end we first remind the reader that at the $k$'th step of the simulation the state vector is on the plane $a+b+c=P_{o}+k$. Thus for this graph instead of providing a three dimensional plot of the velocity field we can present their projections on the plane defined above. Since the sum of the components of the velocity field is normalized to unity everywhere the component along the normal to the plane is the same and thus one must add $\hat{\mathcal{E}}/\sqrt{3}$ to the velocities presented. One word of caution is also in order. The plane where the state vector resides changes at every step of the simulation, so in reading the graphs one must also add this into account to prevent misconceptions such as inferring that the state vector will hit the co-ordinate hyperplanes which is strictly forbidden by the microscopic rules from the outset.

\begin{figure}[t]
\includegraphics[scale=1]{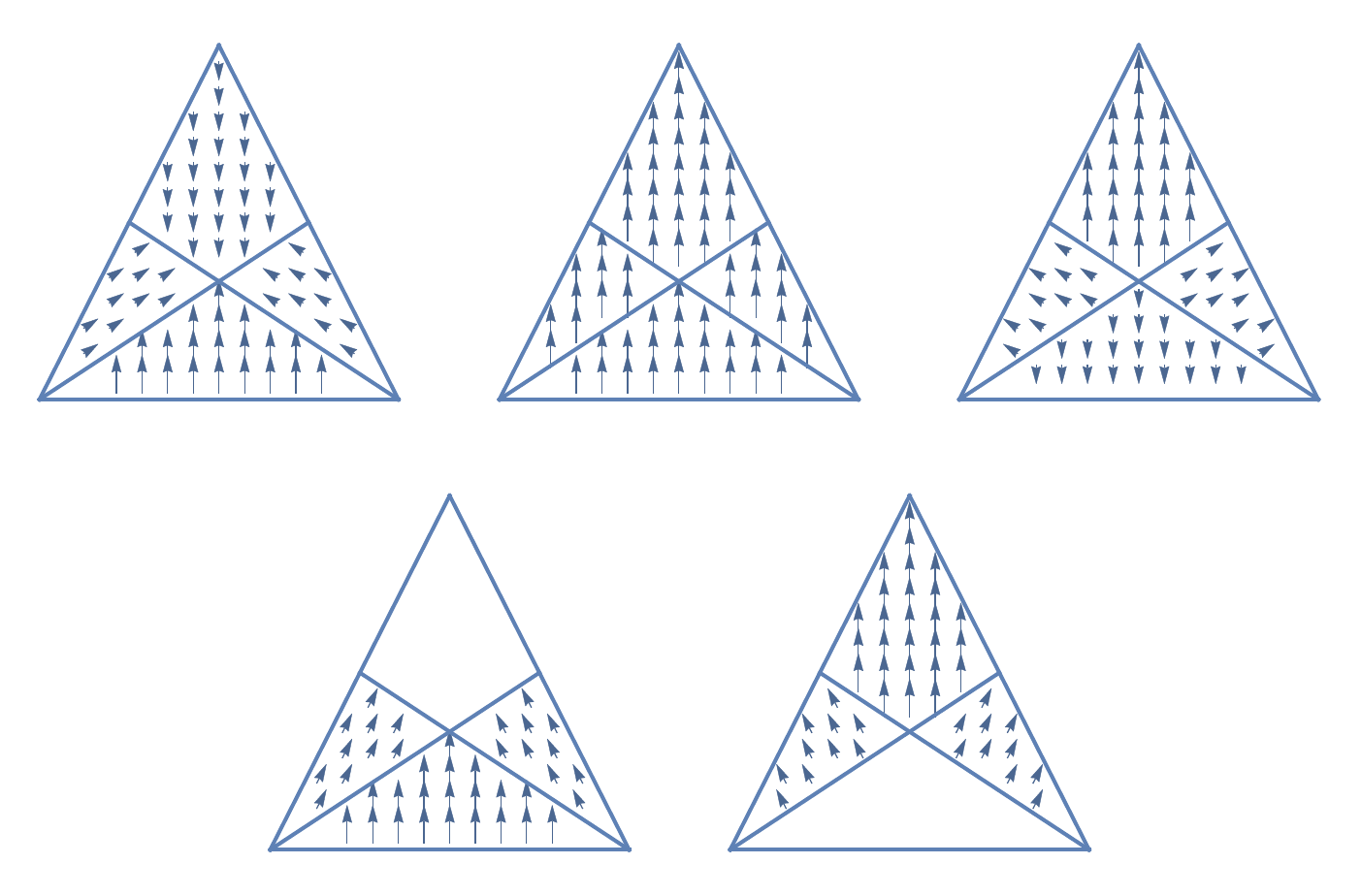}
\caption{The velocity fields for the tree graph with three vertices. The figure represents the projections of the velocity field on the plane $a+b+c={\rm const}$ and thus a perpendicular component of strength $\mathcal{E}/\sqrt{3}$ is to be understood. The top corner of the triangle represents the point $c=k$ and the left bottom corner represents $a=k$. The hyperplanes relevant to the graph at hand are also plotted; the $c=a$ and $c=b$ lines. Also in the naming scheme we have used for the chambers the bulk regions are $S_{T}$, $U_{R}$, $S_{B}$ and $U_{L}$ as read clockwise starting from the top region defined by the hyperplanes. The top row  represents the cases $p=0$, $p=1/2$ and $p=1$ respectively. The bottom row represents $p=1/3$ and $p=2/3$ respectively. Note that for the cases $p=1/3$ ($p=2/3$) the velocity field in the chamber $S_{T}$ ($S_{B}$) do not have any component parallel to the $a+b+c={\rm const}$ plane and  have only vertical components.}\label{fig:meem}
\end{figure}

\subsubsection{Initial data and asymptotics}

\noindent{\bf{Bulk Initial Distribution:}} The first initial data type we shall study is the one where we sample the initial state equally likely from a hypercube or hypersphere; thus the chambers are sampled with probabilities proportional to
their volume ratio. This means that the initial state will be in chamber $S_{T}$, $S_{B}$, $U_{R}$ and $U_{L}$ with probabilities $2/6$, $2/6$, $1/6$ and $1/6$ respectively. Now let us consider the case with $p=1$ for concreteness. As we have seen in this case the velocity field in $U_{R}$ and $U_{L}$ are parallel to the hyperplanes that separate them with $S_{T}$ and hence any state in these chambers that are unstable
will eventually make a transition to the chamber $S_{T}$ and remain there. This means that in the asymptotic limit the probability that the state will be in chambers $S_{T}$, $S_{B}$, $U_{R}$ and $U_{L}$ are given as $2/3$, $1/3$, $0$ and $0$ respectively; a result that we have substantiated with numerical simulations. As another example for this
initial conditions let us consider the case where $1/3<p<2/3$. It is clear from the plots of the velocity field that in this case only $S_{T}$ is a stable chamber and all the initial states will converge here; a fact we have also substansiated with simulations.

\noindent{\bf{Zero Initial Distribution:}} The second type of initial conditions is the one where all agents start with the same point which can be chosen to be zero without loss of generality. To keep the analysis short we shall study this only for the case of $p=1$. The most important difference of this type of initial conditions is that there will always be at least one agent with zero point and hence the state of the system 
travels not in the chambers that constitute the bulk but is always on its edges defined by the co-ordinate hyperplanes. We thus expect rather different behaviour as opposed to the previous case. This is, as we have stressed before, due to the broken ergodicity inherent to the model.

\begin{figure}[t]\label{fig:32bor}
\includegraphics[scale=0.5]{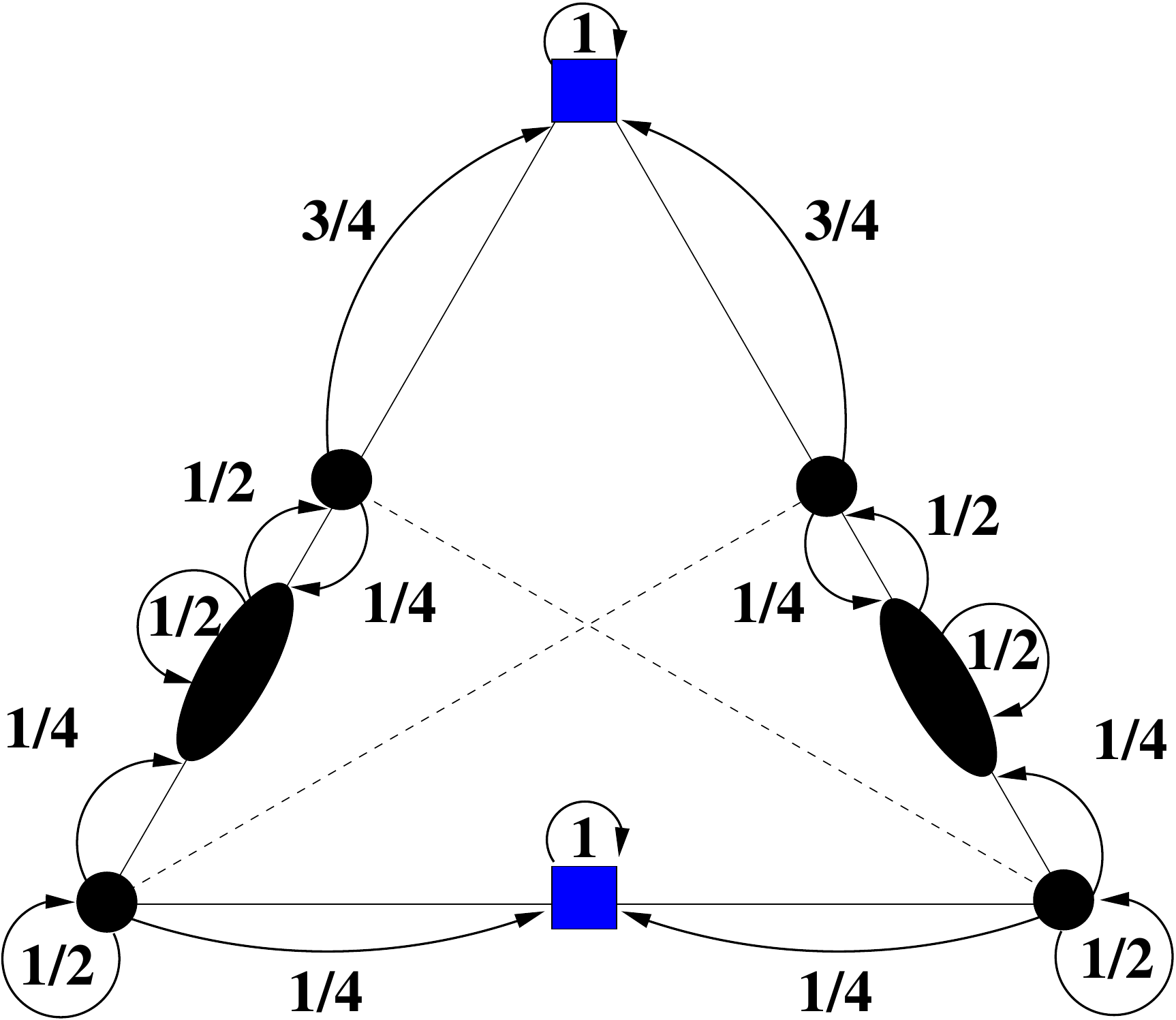}
\caption{The transition probabilities for the tree graph with three vertices for the case when all the agents start at zero point. All the relevant cases are described though the graph can be simplified by aggregating the regions filled black on either side. Note that when the agents start at zero point they are not at the regions shown in the figure; but after the first iteration of the single unit of competition the state will go to the top corner with probability $1/2$ and to the bottom right and left corners with probability $1/4$ respectively. Thus the figure is valid after the first iteration and this has to be taken into account for calculations using the Markov matrix. Note also that the squares filled blue are symbolically representing the regions $S_{T}$ and $S_{B}$: The squre at the bottom represents the entire bottom side of the triangle except its extremities that is the intersection of $S_{B}$ with the co-ordinate hyperplane $c=0$. Similarly, the square at the top represents the entire intersection of $S_{T}$ with the co-ordinate hyperplanes $a=0$ and $b=0$.}
\end{figure}

One can, by direct enumeration, show that if we start every agent with zero points, the asymptotic state will be in the chamber $S_{T}$ with probability $3/4$ and in the chamber $S_{B}$ with probability $1/4$, and it is never in the chambers $U_{L}$ or $U_{R}$. On the other hand one can also study the infinite time behaviour by approaching the system as a Markov process with a transition matrix between relevant locations
of the state vector; which is of course an equivalent method. We present this as a graph rather than as a matrix in the same presentation we used for plotting the velocity field. From the graph one can read out the transition matrix and take its infinite power which will duplicate the results just mentioned. All of these results are substantiated by direct numerical simulations.

\subsection{General properties of the solutions to Eq.\ref{eq:voila}}

In this section we discuss general properties of the speed constraint equation on the configurations. The listing of all possible solutions for all $p$ depends on the particulars of the graph. However the following algorithm will
provide all the possible configurations.

First we remember that for $p=1/2$ all speeds are strictly given as $z_{i}=d_{i}/2E$. For $p>1/2$, remembering that the constraint depends on step functions and hence an ordering of points we must check all the acyclic orientations of the graph. For $p=1$ some of these will provide solutions and some will not and as we run $p$ to lower values some will fail to be solutions an will disappear from the list. Since the constraint equation is linear in $p$ all the curves are linear as in Fig.\ref{fig:n3l21} and they meet with the solutions of $p=1/2$. For $p<1/2$ we look for all orientations including the ones that have equality of neighbouring speeds and even include cyclic orientations which will mandate the equality of speeds of vertices in the cycle; which we have dubbed before as partial orientations. Again some of these will provide solutions
and some will not. For example one can show that for the path graph with 4 vertices the orientation $z_{1}<z_{2}=z_{3}>z_{4}$ will
provide a solution for $1/4\leq p\leq 1/2$. As one lowers $p$ there may be a critical point when no  partial orientations provide any solutions and thus there can be only one solution: All agents travel at the same speed given as $1/N$, which comes from the normalization of the velocity vector.

What is the criterion for the case where all agents travel at the same speed? We have already shown that for regular graphs all agents travel at the same speed for $p\leq 1/2$. But for other cases let us consider the agent that has the minimum number
of neighbours. Clearly these agents have the minimum speed at $p=1/2$ and as one decreases $p$ their speed must increase. If all the agents travel at the same speed below a critical value $p_{c}$ this speed must be given both as $1/N$ and as the speed of
this lowest speed agent which is clearly $q_{c}d_{\rm min}/E$. So we get the condition 

\be
\boxed{1/2 \leq q_{c}=\frac{E}{Nd_{\rm min}}\leq 1}
\ee

What this condition means is that if it is not satisfied from above some agents will still have different speeds even for $p=0$. This behaviour was already hinted by the bucket-pipe analogy we have discussed.

Note that the above condition is  validated for regular graphs where $Nd=2E$ and hence $q_{c}=1/2$. We also realize that for tree graphs where $d_{\min}=1$ and $E=N-1$ all agents travel at the same speed below $q_{c}=(N-1)/N$. This can also be very clearly understood from the bucket-pipe analogy we have mentioned before: The the amount of water disperses equally  on a tree graph for $p=0$. 

For graphs that are not tree yet are still planar we can use the known result $N-E+C=1$ where $C$ is the number of independent cycles in the graph; we shall get

\be
1/2 \leq q_{c}=\frac{N-1+C}{Nd_{\rm min}}\leq 1.
\ee

For instance if $d_{\rm min}=1$ for a planar graph; for $C=0$ all agents travel at the same speed below $q_{C}=(N-1)/N$ as the graph will be a tree; for $C=1$ all agents travel at the same speed at $q_{c}=1$ and for $C>1$ we realize that there will be more than one possible speed at $p=0$. 

If $d_{\rm min}=2$ the graph can not be a tree and if planar must have $C\geq 1$, this saturates the lower bound and we shall
obtain $C<N+1$ as the condition for all agents travelling at the same speed below $q_{c}$.

\section{The ring graph}

The ring graph is a regular planar graph with a single cycle and hence $E=N$. For $p\leq  1/2$ all agents will travel with the same asymptotic speed. As we have stated for regular graphs the statistical properties $p>1/2$ cases are identical and
from now on we shall take $p=1$. Let us remember that in this regime neighbouring agents will have different asymptotic speeds.

The condition on the asymptotic speeds will now read

\[
z_{i}=\frac{1}{N}\left[\theta(z_{i}-z_{i+1})+\theta(z_{i}-z_{i-1})\right],
\]

\noindent Using the scaling invariance of the step functions we shall redefine these speeds so that one will have

\begin{equation}{\label{eq:ringvel}}
\boxed{z_{i}=\theta(z_{i}-z_{i+1})+\theta(z_{i}-z_{i-1})}
\end{equation}

\noindent so that $\sum_{i}z_{i}=N$ and the graph average of the speeds become $1$. This means that all $z_{i}$ are in the set $\left\{2,1,0\right\}$, which we would like to relabel as $\{L,M,S\}$, where $L$, $M$ and $S$ refer to largest, middle and smallest of possible speeds respectively.

Furthermore since no neighbouring agents will have the same $z$ value we can ignore zero argument in the
$\theta$ functions and as a result of this we have the following conditions

\begin{equation}
\theta(z_{i}-z_{i\pm 1})^{a}=\theta(z_{i}-z_{i\pm 1}),
\end{equation}

\noindent for any $a>0$.

\subsection{Constraints on the configurations}

So the model of two agent games on the ring looks like  a 3-colouring problem. This is a well known system and is directly tied to the 3-state anti-ferromagnetic Potts model on the ring graph at zero temperature. However this analogy is altered by constraints that naturally arise form the two-agent model via Eq.\ref{eq:ringvel} and by the broken ergodicity inherent in the model.

For instance consider the following configuration

\[
\cdots SMS\cdots
\]

This does not make any sense since the agent in the middle will surely win all games against its neighbours. Assuming the rest of the configuration is meaningful in this respect, this configuration should evolve to

\[
\cdots SLS \cdots
\]

Via a similar reasoning one has

\[
\cdots LML
\cdots\longrightarrow\cdots LSL\cdots
\]

Thus the configuration space of our  model actually is that of a 3-state anti-ferromagnetic Potts model at zero temperature where the colors are labeled as $\{ L,M,S \}$ and where the tilings $SMS$ and $LML$ are forbidden. These conditions are symmetric with respect to $L\leftrightarrow S$ and thus the full symmetry group of the configurations for a given $N$ are $G=D_{N}\times C_{2}$ as opposed to $D_{N}\times S_{3}$ which binds the unconstrained three colourings of the cycle graph. Here $C$, $D$ and $S$ mean the cyclic, dihedral and permutation groups respectively.

As the constraint equation on the speeds involve the orientations of the graph we can also see the imposibility of a tile like $LML$ or $SMS$ as follows: M-type agents mean one has $\cdots >M>\cdots$ type of ordering. Similarly L-type agents have $\cdots <L>\cdots$ and S-type agents have $\cdots >S<\cdots$ and thus no orientation can provide the mentioned tiles.

To express these observations in mathematical terms, let us define the following quantities

\begin{subequations}
\bea
L_{i}&\equiv & \theta(z_{i}-z_{i+1})\theta(z_{i}-z_{i-1}),\\
S_{i}&\equiv &\theta(z_{i+1}-z_{i})\theta(z_{i-1}-z_{i}),\\
M_{i}&\equiv &\theta(z_{i}-z_{i+1})\theta(z_{i-1}-z_{i})+\theta(z_{i}-z_{i-1})\theta(z_{i+1}-z_{i}).
\eea
\end{subequations}
which are simply the eigenfunctions of $z_{i}$ in that they obey

\begin{subequations}
\begin{eqnarray}
z_{i}L_{i}&=&2L_{i},\\
z_{i}M_{i}&=&M_{i},\\
z_{i}S_{i}&=&0.
\end{eqnarray}
\end{subequations}

These variables further satisfy $L_{i}^{2}=L_{i}$, $M_{i}^{2}=M_{i}$, $S_{i}^{2}=S_{i}$ and $S_{i}M_{i}=S_{i}L_{i}=M_{i}L_{i}=0$. Since they are exclusive one also has $L_{i}+M_{i}+S_{i}=1$. Furthermore 

\be
L_{i}L_{i+1}=M_{i}M_{i+1}=S_{i}S_{i+1}=0,
\ee
which simply re-express that this is a colouring problem.

Now it is easy to show that 
\begin{subequations}
\bea
S_{i-1}M_{i}S_{i+1}&=&0, \\
L_{i-1}M_{i}L_{i+1}&=&0,
\eea
\end{subequations}
as the mathematical expressions of the constraint on the configurations.

From the constraints it is clear that the M-type agents are residing over a checkerboard formed by $L$ and $S$ type agents. Note that there will be no alteration on the checkerboard in view of the constraint mentioned above. Since for an alteration to occur
a tile of the forbidden type must be used. As an example this is a valid tiling where there is no alteration of the
checkerboard background $\cdots LSLSMLSMLMSMLSLSLSLSMLSLM\cdots$. 

Furthermore once a given configuration is found the only affect of $L\leftrightarrow S$ is to yield just a single other configuration. From this perspective one can, when convenient, see the background only as a single type of object say $o$ and the above configuration will become $\cdots ooooMooMoMoMooooooooMoooM\cdots$. Also, as a consequence of these constraints if the ring has even(odd) number of vertices there can only be an even(odd) number of M-type vertices in any configuration.

On the other hand it is also fruitful to remember the connection to hyperplane arrangements and hence to the generic vertex-model character of the system we study. Any valid configuration described above can be mapped one-to-one to an acyclic orientation on the ring graph and it must be a stable orientation from previous arguments we have presented. Not all acyclic orientations however are
goind to yield a colouring commensurable with the constraints. Yet we have already shown that the exit from an unstable chamber
rests on first passage times to the neighbouring stable chambers in the configuration space. For the ring graph one can show that
the unstable chambers are those that contain at least three similar  signs in a row. For instance the following acyclic orientation
$\cdots <><<><<<<>>>>><><\cdots$ will be mapped to an unacceptable colouring in the form $\cdots SLMSLMMMSMMMMLSL \cdots$. As we have discussed before, the resolution of this will be via the first passage properties of the series of like arrow regions in the orientation. But if we sample equally likely from the acyclic orientations the splitting probabilities will be evenly distributed and thus we simply flip orientation of the links in the chain randomly. Note that the change of the orientation of a single edge in a series of like arrows will remove two M-type agents and will replace it with a checkerboard element -LS or SL- congruent with the rules. Thus if we sample equally likely from acyclic orientations the dynamics is such that given a $k$ chain of like signs one performs sign flips in a region of length $k-2$ and this is like dimer adsorbtion in a chain of length $k-1$. However, choosing initial conditions equally likely from acycic orientations is a bit unnatural for the microscopic model we have. The most natural ones are either starting all from zero
points or sampling equally likely from inside a large cube. The latter will sample from acyclic orientations commensurable
with their relative volumes. We shall come back to these issues in a moment.

Going back to the study of constraints the following relations are easily obtained 
\be
z_{i}=2L_{i}+M_{i},
\ee
along with 
\be
z_{i}^{2}=z_{i}+2L_{i},
\ee
and one can iterate this last property to find
\be\label{eq:zscale}
z_{i}^{k}=z_{i}+\left(2^{k}-2\right)L_{i}\;.
\ee
which simply means that the probability distribution of a single speed variable $z_{i}$ depend on a single parameter. 

Last of all we would like to talk about an exact chain of relations between expectation values of the
different tilings. As well known these are obtained by first starting with a single site's configurations: $L_{i}$, $S_{i}$ or $M_{i}$. At the zeroth level we have from the probability conservation the following

\be
\langle M\rangle+2\langle L\rangle =1,
\ee
where we have used the symmetry group $G$ to reduce the number of unkowns.

Now by multiplying by either $1_{i\pm 1}=L_{i\pm 1}+S_{i\pm 1}+M_{i\pm 1}$ and using all the $G$ symmetry
we shall arrive at the following set 

\bea
2\langle ML\rangle &=& \langle M\rangle, \\
\langle LS\rangle &=& \langle L\rangle -\langle ML\rangle.
\eea
Thus if $\langle M\rangle$ is known so are the two tilings. However if we proceed further, the symmetry group $G$ is not enough to reduce the unknowns to the level of the distinct equations one
can obtain from this procedure and one will have to introduce at each level new parameters. One can show that these can 
be chosen to be $\langle M\rangle$, $\langle Mo M\rangle$, $\langle Moo M\rangle$, and so on. Clearly this amounts to giving the chemical potential and interaction energy if one views the $M$ type agents as particles of a gas over an $LS$ checkerboard background.

\subsection{The speed distribution of agents in the mean}

An important property of the model is the number of agents having points less than say $s$. This information can be obtained by studying the following random variable

\be
N_{s}(\vec{X})=\sum_{i}\theta(s-X_{i}).
\ee

One can use the representation 

\[
\theta(u)=\lim_{\alpha\to\infty}\left(1+\tanh(\alpha u)\right)/2,
\]

\noindent perform a series expansion, swithch to the scaling ansatz $z_{i}=x_{i}/t$ and use Eq.\ref{eq:zscale} to arrive at

\be
\frac{1}{N}\langle N_{z}\rangle=n_{L}\theta(z)+(1-2n_{L})\theta(z-1)+n_{L}\theta(z-2),
\ee
where we have defined $z=s/t$.
Thus we see that the fraction  of agents with speed $0$ is equal to the fraction of agents with speed $2$; somewhat expected from the $L\leftrightarrow S$ symmetry of the constraints we have found. There remains to find $n_{L}$ or $n_{M}$ for that matter since $n_{M}+2n_{L}=1$.

\subsubsection{Various Initial Conditions and Simulation Results}

\noindent{\bf Starting from an almost cyclic orientation:} Here the initial condition is such that the points of the agents
are large and follow the ordering given by the orientation $\cdots>><>>>\cdots$, that is we have $N-1$ same side orientations and one opposite. If the initial points are distributed evenly inside the chamber defined by this orientation the splitting probabilities to the neighbouring chambers are evenly distributed as well. The problem therefor is that of a dimer adsorption
in a path graph of length $N-1$. In the limit of $N\to\infty$ we shall therefore get \cite{flory,redbook2}
\[
n_{M}=e^{-2}\approx 0.1353.
\]
which is in full accord with our simulations.

\noindent{\bf Starting from acyclic orientations sampled equally:} Here we simply sample from acyclic orientations equally likely.
A given acyclic orientation has a known distribution $p_{k}$ of k consecutive like signs, which in the limit $N\to\infty$ have a simple expression. Since there will be various chains of like signs, the eventual ratio of $M$-type agents will be given as

\be
n_{M}=\frac{1}{N}\sum_{k=2}^{N-1}g(k-1)p_{k},
\ee
where $g(k)$ is the average number of unoccupied sites, which are M-type agents in our context, in the dimer adsorbtion problem.
They are given as \cite{redbook2}

\be
g(k)=k-\frac{1}{(k-1)!}\frac{d^{k-1}}{dz^{k-1}}A(z)\vert_{z=0},
\ee
with 
\be
A(z)=\frac{1-e^{-2z}}{(1-z)^{2}}.
\ee

Also in the large $N$ limit $p_{k}$ is  given as

\be
p_{k}=(N-k+1)\left(\frac{1}{2}\right)^{k+1}.
\ee

By performing the sum above with Mathematica for $N=1000$ we find

\be
n_{M}=0.183(1).
\ee
which is in full accord with our simulations.

\noindent{\bf Bulk type initial conditions:} This is the first of a realistic initial condition we consider: We sample points form a large $N$ dimensional cube, which in turn yields an acyclic orientation sampled by its relative volume. The
relative volumes of chambers are easy to define but we do not know of any closed form formula for the ring graph. However the
ratio of $M$ type agents can be expressed using the result on equal sampling of acyclic orientations modulated by the mentioned volume ratios and for small rings an exact calculation is possible. The simulation results are

\be
n_{M}=0.198(1).
\ee

\noindent{\bf Zero initial condition:} In this case all agents start with the same point which, in view of the mentioned shift symmetry of the velocity field, can be taken to be zero without loss of generality. The simulation results are

\be
n_{M}=0.175(1).
\ee

\subsection{Two meaningful approximations}

In this subsection we shall provide two approximations schemes. One for the estimation of the ratio of number of agents having  types of speeds pertinent to the graph and one for
the statistics of the configurations. 

\subsubsection{Congruent minimal tiles approximation for the density of types of vertices}

One can show that the following exact relations hold 

\begin{subequations}
\bea
M_{i}&=&L_{i-1}M_{i}S_{i+1}+S_{i-1}M_{i}L_{i+1},\label{eq:sugar}\\
L_{i}&=&S_{i-1}L_{i}S_{i+1}+M_{i-1}L_{i}M_{i+1}+M_{i-1}L_{i}S_{i+1}+S_{i-1}L_{i}M_{i+1},\\
S_{i}&=&L_{i-1}S_{i}L_{i+1}+M_{i-1}S_{i}M_{i+1}+M_{i-1}S_{i}L_{i+1}+L_{i-1}S_{i}M_{i+1}.
\eea
\end{subequations}

Note that on the first equation there are an equal number of M-type vertices on both sides of the equation. Similarly on the second(third) equation there are an equal number of L(S)-type vertices on both sides. Furthermore the 3-tiles described above
are the smallest tiles that are congruent with all the constraints of the system. As this equation is exact it will also hold for the configuration averages of the tiles that takes part in it. 

The approximation is to assume that the configuration averages of the 1-tiles on the left are proportional to the number of 3-tiles to the right. This does not mean that we expect all the 3-tiles on the right hand sides occur equally likely, this does not happen as we have checked in simulations. We simply assume that the sums contrive to yield a number proportional to the number of 3-tiles on the right hand sides. The normalization of the probability does the rest and we have

\be
\left.\begin{array}{ccc}
n_{L}=n_{S} &=& 4/10 \\ 
n_{M}&=&2/10
\end{array}\right\}\;\;\;\;\;{\rm CMT\;\;\;approximation}
\ee

Note that this approximation is quite good for the bulk type initial conditions and still close for zero initial conditions. We shall see that this scheme provides a rather adequate estimate for higher dimensional tori as well.

\subsubsection{Chemical potential approximation for the statistics of the configurations}

Due to the broken ergodicity inherent in the model not all configurations occur with the same probability. As we have previously
mentioned one can in principle calculate them from the transition probabilities from unstable chambers to the stable ones. However
this is a formidable task as we have laid out for the tree graph with three vertices and we do not yet have an approach that is
sure to give a closed form result. 

However it is also clear that the most important quantity is the ratio of M-type agents. The chemical potential approach we advocate is to
assume that all configurations with a given number of M-type agents occur equally likely weighed with a chemical potential $\mu$. This potential describes the cost of removing a pair of LS agents from the full checkerboard configuration of unspecified absolute
probability and replace them with two M-type agents randomly distributed over the background such that they are not next to each
other.

For instance in this approach the ratio of M-type agents will be given as

\begin{equation}
n_{M}=\frac{1}{N} \frac{2\sum_{k=0}^{N/2}\;2k\;\mu^{2k}\binom{N-2k}{2k} }{2\sum_{k=0}^{N/2}\;\mu^{2k}\binom{N-2k}{2k}},
\end{equation}
where for simplicity we assumed that $N$ is even and the factors of $2$ in the numerator and the denominator are simply there because of $L\leftrightarrow S$ symmetry; as we see they cancel.

The expression above has a well defined large $N$ limit given as

\be
n_{M}=\frac{1}{2}\left(1-\frac{1}{\sqrt{1+4\mu}}\right).
\ee

What one can do now is to associate a particular value of $\mu$ for a given type of initial condition, from the data or from the CMT approximation. The rest will be to calculate various statistical quantities from the partition function. In the appendix  we give full details of this approach.

Of course this is only an approximation and can be further ameliorated at the next level by the inclusion of a new separate chemical potential for tiles of the type $MoM$ where $o$ stands for L or S-type vertices. However we do not pursue it further here; the chemical potential approach alone provides a rather good approximation.

\subsection{Separation distribution of two M type agents}

One important quantity we can study is the distance distribution between two M-type agents. We know that the
space between them is filled with an island of checkerboard configuration of L and S type agents. Since we are studying the ring a configuration, $N_{M}$ M-type agents will give rise to $N_{M}$ checkerboard islands. Since the total number of agents is conserved we arrive at

\be
N=N_{M}+\sum_{i=1}^{N_{M}}k_{i}.
\ee

Since when we let $N\to\infty$ so will $N_{M}$, dividing by $N_{M}$ we arrive at an exact result

\be
\langle k\rangle=\frac{2n_{L}}{n_{M}}=\frac{2n_{L}}{1-2n_{L}}.
\ee

To estimate the distribution using the chemical potential approach we first recall that in this approach configurations with a given number of M-type agents are sampled equally likely. This means the following: Let us assume we walk on a given configuration given that we start on an M-type agent and count the number of sites to reach the next M-type agent. The mean number of these agents is given as $n_{M}$ and since in the chemical potential approximation the configurations do not care where the M-type particles reside, we expect this to be a Poisson process. As this is a conditional probability we thus have for distribution of distances

\be
P(k)=n_{m}(e^{\alpha}-1)e^{-\alpha k},\;\;\;\;\;\;\;\;\;{\rm with}\;\;k\geq 1\;.
\ee

Calculating $\langle\delta\rangle$ with this distribution gives us the value of $\alpha$

\be
\alpha=\ln\left(\frac{2n_{L}}{4n_{L}-1}\right).
\ee

We derive this expected result in  the Appendix. The comparison with actual data and this estimate is presented in Fig.\ref{fig:distring}.

\begin{figure}[t]
\includegraphics[scale=0.5]{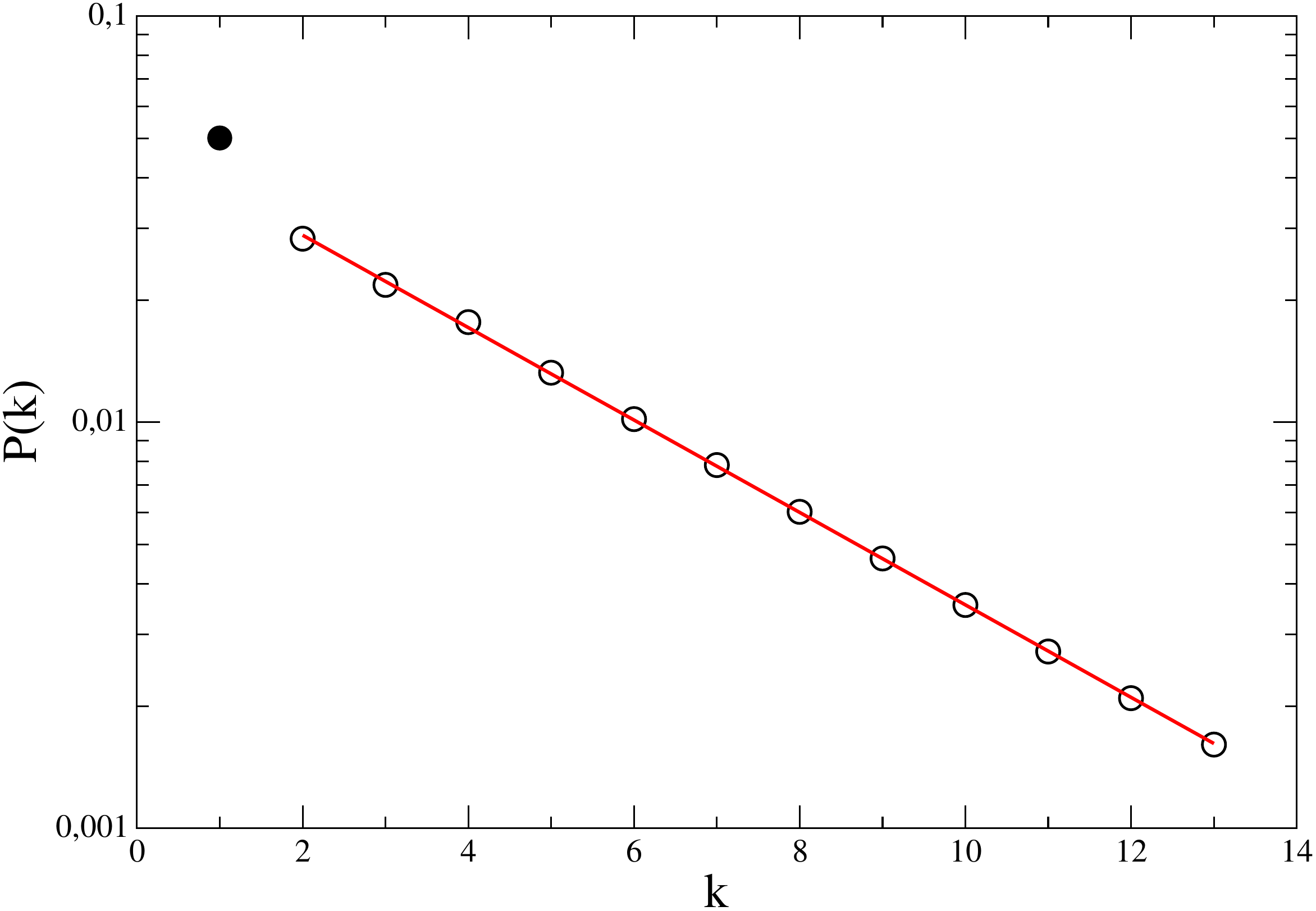}
\caption{The probability distribution of the distance between two M-type agents for the model on the
cycle graph. The data is obtained for bulk type initial conditions. Note that for $k=1$ the description of the chemical potential approach fails to provide adequate description for tiles of type $MoM$ and this can be remedied by introducing a new chemical potential for this tile.  This failure is milder for say zero initial conditions since $n_{M}$ is smaller and hence $MoM$ type tiles are less common. However note that for $k\geq 2$ the chemical potential approach is rather good.}\label{fig:distring}
\end{figure}

\subsection{The pair correlation function of asymptotic speeds}

Another quantity of interest is the pair correlation function
\be
c_{k}\equiv\langle z_{i}z_{i+k}\rangle=4\langle L_{i}L_{i+k}\rangle+2\langle M_{i}L_{i+k}\rangle+2\langle L_{i}M_{i+1}\rangle+\langle M_{i}M_{i+k}\rangle,
\ee
where we have used the $G$ symmetry to reduce the form of the right hand side. Note that since $\langle z_{i}\rangle=\langle z_{i+k}\rangle=1$, to obtain the pair cumulant one has to simply subtract 1 from the above equation.

For $k=0$ we use the previous result that $z_{i}^{2}=z_{i}+2L_{i}$ and hence one has
\be
c_{0}=1+2n_{L}\;.
\ee

Furthermore for $k=1$ we have to calculate $\langle z_{i}z_{i+1}\rangle=2\langle L_{i}M_{i+1}\rangle+2\langle M_{i}L_{i+1}\rangle$. But from our constraints $L_{i}M_{i+1}=L_{i}M_{i+1}S_{i+2}$ and
$M_{i}L_{i+1}=S_{i-1}M_{i}L_{i+1}$. Remembering Eq.\ref{eq:sugar} we see that 

\be
c_{1}=2 n_{M}\;.
\ee

\begin{figure}[t]
\includegraphics[scale=0.5]{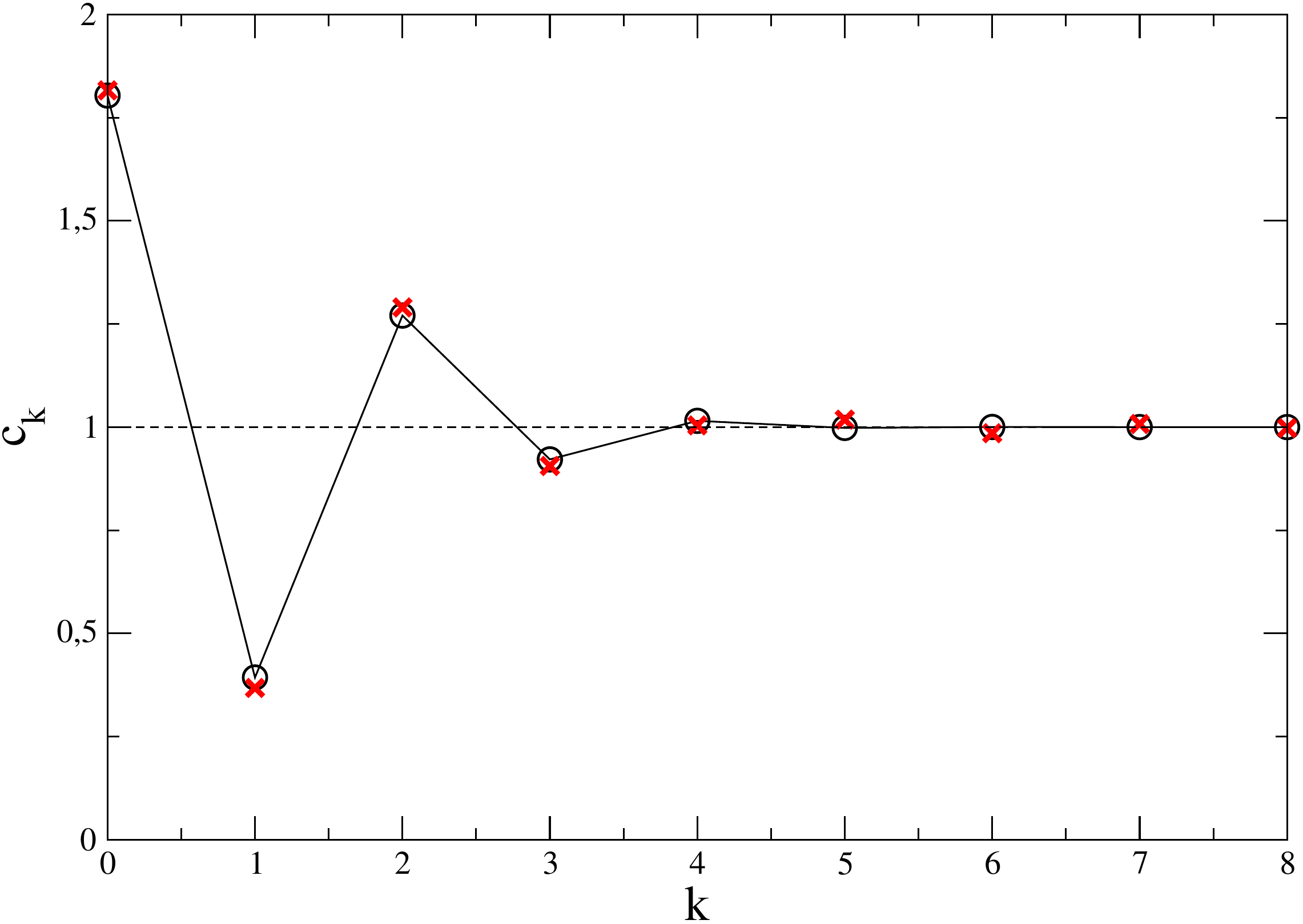}
\caption{The pair correlation function of asymptotic speeds in the ring graph. The data is obtained for bulk type initial conditions. The red crosses are
theoretical calculations; the first two are exact and the others are from the chemical potential approximation which also gives the same result for the first two.}\label{fig:paircorre}
\end{figure}

In the Appendix where we solve the problem with the chemical potential approach we derive an exact expression for the pair correlation.The results of the simulation and these theoretical calculations are shown in Fig.\ref{fig:paircorre}.

\section{The torus graph}

The rectangular lattice with periodic boundary conditions is topologically equivalent to a torus. Since all
agents have the same number of neighbours this graph is regular. We consider the case where links are
placed along the orthogonal directions so every agent has 4 neighbours. Since the graph is regular,
for $p\leq 1/2$ all agents have the same asymptotic speed and all cases $p>1/2$ are equivalent to the case
$p=1$. For $p=1$ the constraint of the speeds are given, with an obvious normalization similar to the one we
have done for the ring graph, as

\be 
\boxed{z_{i,j}=\theta(z_{i,j}-z_{i+1,j})+\theta(z_{i,j}-z_{i-1,j})+\theta(z_{i,j}-z_{i,j+1})+\theta(z_{i,j}-z_{i,j-1})}
\ee

So the speeds are in the set $\{0,1,2,3,4\}$. Let us denote these speeds as $\{S,M^{-},M^{0},M^{+},L\}$ and 
the mean fraction of agents having those as $n_{S}$, $n_{M^{-}}$, $n_{M^{0}}$, $n_{M^{+}}$ and $n_{L}$ respectively.

In analogy to the ring graph case we can introduce the variables $S_{i,j}$,$M^{-}_{i,j}$, $M^{0}_{i,j}$, $M^{+}_{i,j}$ and $L_{i,j}$ which will be the eigenvectors of $z_{i.j}$. Let us collectively call them $\sigma^{\alpha}_{i,j}$ with $\alpha\in\{0,1,2,3,4\}$. These, as before, list all possible speeds of a given agent in terms of all the 5-node
tilings centered about a given agent that are commensurable with the velocity equation. Similarly one can also easily show that
$(\sigma^{\alpha}_{i,j})^{2}=\sigma^{\alpha}_{i,j}$ and $\sigma^{\alpha}_{i,j}\sigma^{\beta}_{i',j'}=\delta^{\alpha\beta}\sigma^{\alpha}_{i,j}$ if $(i',j')$ denote a neighbour of the node $(i,j)$. And thus we see that the phase space of the torus model is that of an anti-ferromagnetic Potts model with 5 states at zero temperature, and hence 5-colourings, constrained with the velocity equation.

One can show that constraints similar to the ring graph case can also be constructed for the torus graph via these variables. Here instead of presenting the full list we simply write down how they will emerge.

\begin{subequations}
\bea
\sigma^{\alpha}_{i,j}&=&1_{i-1,j}1_{i+1,j}1_{i,j+1}1_{i,j-1}\sigma^{\alpha}_{i,j}.\\
\eea
\end{subequations}

To the right of the above equation we will have all the 5-tiles colourings commensurable with the velocity equation. Since this equation is exact, it will also hold in the mean and will constitute the basis for the CMT approximation as in the ring diagram.

\subsection{The speed distribution of agents in the mean}

In precise analogy to the ring graph, one can, after some tedious but straightforward calculation, show that the mean number of  agents having speeds less than a given value $z$ obeys the following

\be
\frac{1}{N}\langle N_{z}\rangle=n_{S}\theta(z)+n_{M^{-}}\theta(z-1)+n_{M^{0}}\theta(z-2)+n_{M^{+}}\theta(z-3)+n_{L}\theta(z-4),
\ee

along with 

\begin{subequations}
\bea
n_{S}&=&n_{L},\\
n_{M^{-}}&=&n_{M^{+}},\\
n_{M^{0}}+2n_{M^{+}}+2n_{L}&=&1.
\eea
\end{subequations}

\subsubsection{Simulation Results}

\noindent{\bf Bulk type initial conditions:} Here we sample initial conditions equally likely from an N-dimensional cube. The size of the cube is large enough so as to discriminate different chambers in the configuration space. The simulation results are

\begin{subequations}
\bea
n_{S}=n_{L}&=&0.310(1),\\
n_{M^{\pm}}&=&0.133(1),\\
n_{M^{0}}&=&0.112(1).\\
\eea
\end{subequations}

\noindent{\bf Zero initial condition:} Here we start all agents from zero points. The simulation results are

\begin{subequations}
\bea
n_{S}=n_{L}&=&0.336(1),\\
n_{M^{\pm}}&=&0.116(1),\\
n_{M^{0}}&=&0.096(1).\\
\eea
\end{subequations}

\subsubsection{CMT approximation}

As in the ring case the CMT approximation consists of counting all the 5-tiles commensurable with constraints and assuming that the vertex type in the middle has a configuration average proportional to the number of 5-tiles that has that type of agent in the middle. The counting yields the following

\begin{subequations}
\bea
n_{S}=n_{L}&=&\frac{256}{824}\approx 0.3107, \\
n_{M^{\pm}}&=&\frac{108}{824}\approx 0.1311,\\
n_{M^{0}}&=&\frac{96}{824}\approx 0.1165. \\
\eea
\end{subequations}

We see that, as was the case for the ring graph, CMT approximation is very close to the results for bulk type initial conditions.

\subsection{The geometry of the configurations}

\begin{figure}[!tbp]
  \centering
  \begin{minipage}[b]{0.4\textwidth}
    \includegraphics[width=\textwidth]{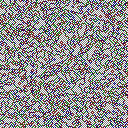}
  \end{minipage}
  \hspace*{1cm}
  \begin{minipage}[b]{0.4\textwidth}
    \includegraphics[width=\textwidth]{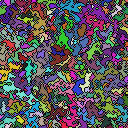}
  \end{minipage}\caption{A sample configuration on the 128x128 torus obtained via zero initial conditions. The picture on the left is where agents of type $L$,$S$,$M^{0}$, $M^{+}$ and $M^{-}$ are represented by grey, white, red, green and blue pixels respectively. To obtain the picture on the right we first coloured all $M$ type agents by black and randomly coloured all $LS$ checkerboard islands for visualization of the cluster shapes. Mild image smoothing is applied for visibility.}\label{fig:conf128}
\end{figure}

A sample configuration on a torus is shown in Fig.\ref{fig:conf128} We note that as in the one dimensional case, there are LS checkerboard islands separated by M-type vertices. There are various geometrical properties one can study. We focus on the distribution of
the areas and of perimeters of LS checkerboard islands. To that end let us assume that we have $N_{c}$ islands in a given
configuration. One has

\be
\frac{1}{N_{c}}\sum_{i=1}^{N_{c}}R_{i}=N_{S}+N{L}=2N_{L},
\ee
and 
\be
\frac{1}{N_{c}}\sum_{i=1}^{N_{c}}{\partial R}_{i}=2(N-2 N_{L}),
\ee
where $R_{i}$ is the size of an island and $\partial R_{i}$ the size of its boundary. The factor of 2 in the second equation above comes from the fact that when one sums  the boundaries consisting of the M-type agents over all islands, boundaries eventually will be counted twice.

We thus immediately see the following

\be
\langle R\rangle=\langle\partial R\rangle \frac{n_{L}}{1-2n_{L}}.
\ee

Note that for zero initial conditions one has $n_{L}\approx 1/3$ and thus $\langle R\rangle \approx \langle \partial R\rangle$. So for this case, in the mean, the checkerboard islands have the same discrete  area and boundary size. Of course such regions have
a wide spectrum of shapes. 

We have also studied the size distribution of the island areas and their perimeters. The results are shown in Fig.\ref{fig:2dscaling}. We see clear scaling behaviour for both the island areas and the perimeter. The scaling exponents are extracted as 
approximately $-1.32$ for the areas and $-1.55$ for the perimeters.

One might have imagined that one could have formed an approach similar to the chemical potential approximation  we have presented for the ring graph. The basis of that approximation was of course is exact tractability. Here in the torus graph our system can
be mapped to a 16-vertex model but does not admit a solution for the Yang-Baxter equation and thus it is not integrable. So we do not pursue such an avenue here.

\begin{figure}[t]
\includegraphics[scale=0.5]{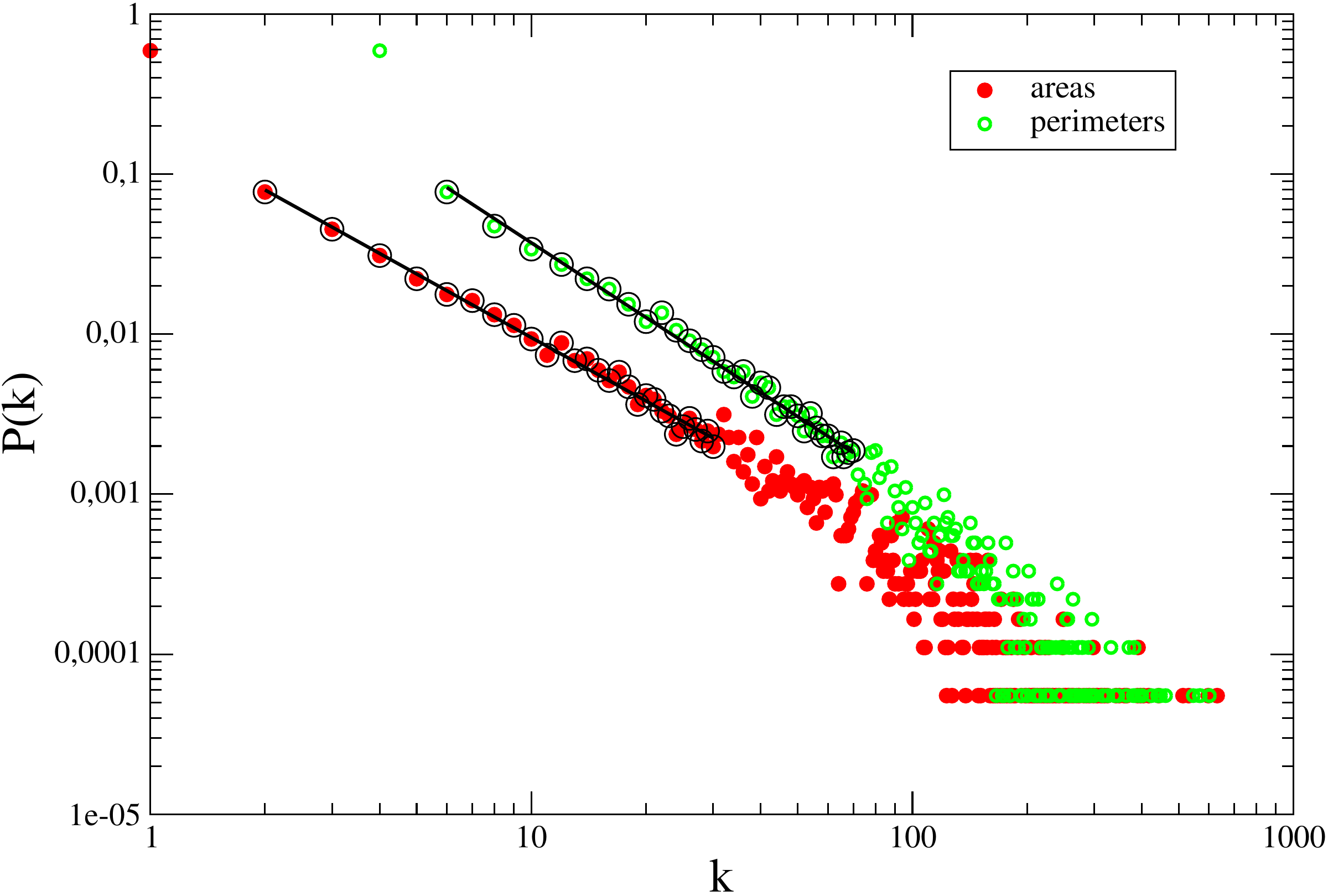}
\caption{Area and perimeter distributions for the LS checkerboard islands. Data is taken from a single 512x512 torus obtained via zero initial conditions. The exponents are approximately $-1.32$ for the areas and $-1.55$ for the perimeters.}\label{fig:2dscaling}
\end{figure}

\section{Digression on d-dimensional tori}

It is clear that our model on d-dimensional tori will be that of $2d+1$ colourings of the graph subject to the constraint of the velocity equation. Here we present results for the mean asymptotic speed distribution of agents in the CMT approximation. The counting consists of  $2d+1$ tilings 

\be
n_{k}=\frac{s(k,d)}{g(d)},
\ee

\noindent where $n_{k}$ is the fraction of agents having speed $k$ where $k\in\{0,1,\cdots,2d\}$. The functions $s(k,d)$ and $g(d)$ are given as follows

\begin{subequations}
\begin{eqnarray}
s(k,d)&=&\binom{2d}{k}k^{k}\left(2d-k\right)^{2d -k},\\
g(d)&=&\sum_{k=0}^{2d}s(k,d),
\end{eqnarray}
\end{subequations}

\noindent where $0^{0}=1$ to be used with impunity. An analysis using the Stirling's approximation yields for instance 

\be
\lim_{d\to\infty}
\;n_{0}=n_{2d}\to\frac{1}{\sqrt{\pi\; (d+1)}}.
\ee

The term $+1$ next to $d$ in the above formula is a phenomenological amelioration of the large $d$ expansion which succeeds to yield rather close results for all $d$, including $d=1$.

Since we know $n_{k}$ from it we can calculate the distribution of agents with speeds less than $z$ where the maximum speed is normalized to $1$. This function is given as

\be
F(z)=\sum_{k=0}^{2d}n(k)\theta(z-\frac{k}{2d}).
\ee

Using Stirling's approximation and integral approximations to the sums we obtain the following

\be{\label{eq:CMTinf}}
F(z)_{d=\infty}=\frac{2}{\pi}\sin^{-1}(\sqrt{z}).
\ee

It is very interesting to recover the famous lead probability distribution of the 1D random walker here. However since the meaning of $F(z)$ is basically of the same nature as the lead probabilities this result is, though maybe not outright expected, understandable. In Fig.\ref{fig:cmt} we present various plots pertinent to the CMT approximation.

\begin{figure}[t]
\includegraphics[scale=0.5]{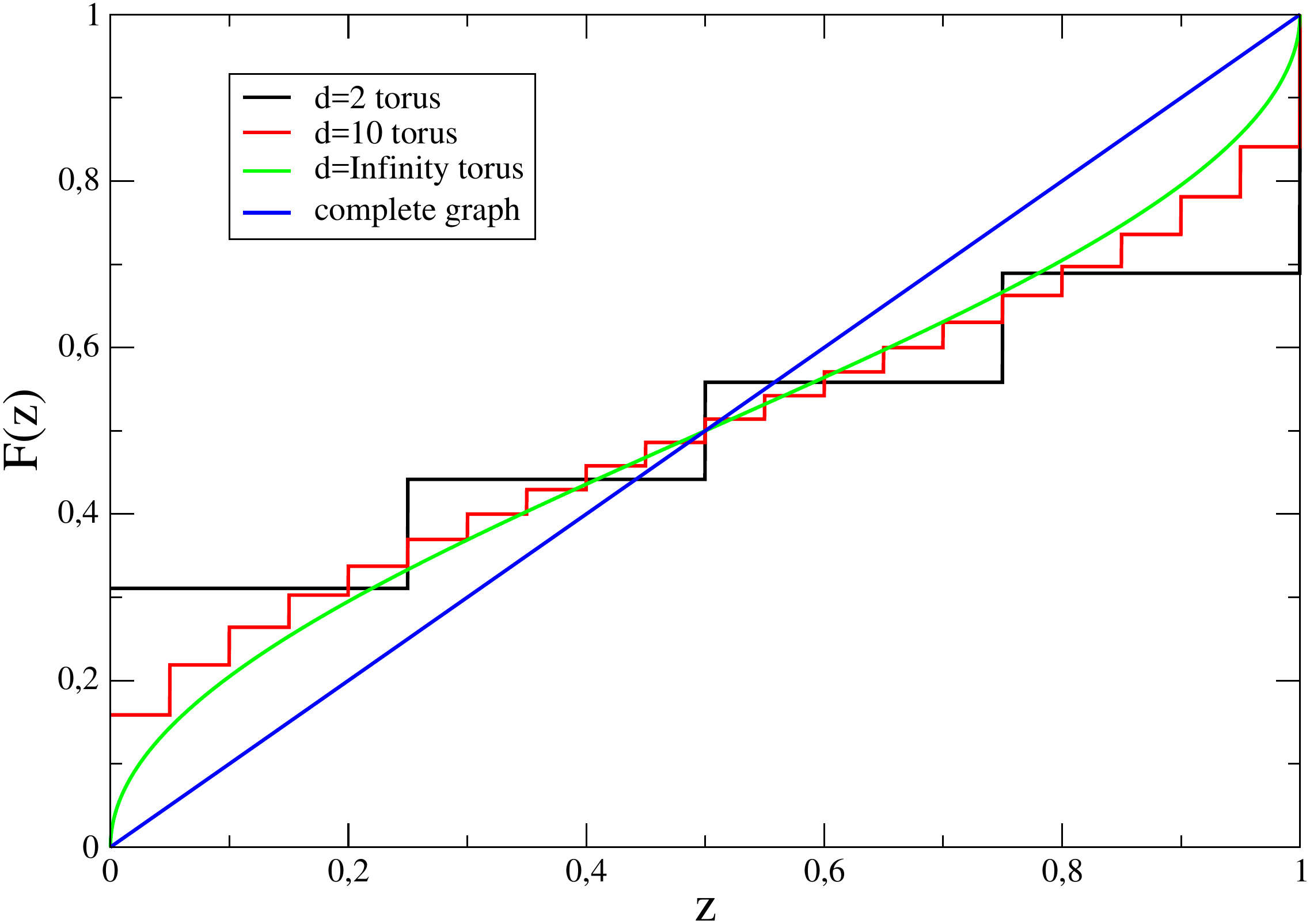}
\caption{The plots of $F(z)$ which is the number of agents with speed less than $z$ in the CMT approximation. The top speed is normalized to $1$ and hence the configuration mean speed is $1/2$. We present plots for $d=2$, $d=10$ and $d=\infty$ along with the well known result for the complete graph which is given as $F(z)=z$. We see that CMT approximation captures the details rather well. Remember that the result for the complete graph is akin to mean field approximation.}\label{fig:cmt}
\end{figure}

It is a valid question up to what dimensionality this pattern of checkerboard islands will be appreciably present. It is clear that
as $d$ increases so does the fraction of L and S type agents. One can come up with a critical dimensionality by assuming that $n^{c}_{L}=1/4$ which would mean that the M type agents overall has reached half of the population. Using the CMT approximation we 
can calculate that for $d=3$ one has $n_{L}\approx 0.265$ and for $d=4$ one has $n_{L}=0.235$. So it seems the critical dimensionality is $4$. Motivated by this estimate we have performed an analysis on the 3D torus with zero initial conditions. The tentative results are in aggrement with the CMT approximation.

\section{Conclusions and future directions}

As we have so far seen two-agent games on graphs can yield a kaleidoscope of interesting phenomena. This richness we believe is due
to its simplicity. The reason for this simplicity lies in the fact that the resolution of the two-agent unit of competition depends
only on step functions. However even if one generalizes the rules of engagement the basic properties we have presented, will, we believe, remain intact. To back this assesment up let us consider two agents with points $x$ and $y$. The probability that the agent with point $x$ will win is

\be
P_{\rm w}=p\theta(x-y)+q\theta(y-x)
\ee
which is a function that starts out at value $q$ and makes a jump to value $p$. Now let us assume we enlarge the region of transition; that is let us take a function that starts out at $q$ and gradually increases to value $p$. The impact of this
generalization will only be to somewhat enlarge the thicknesses of the hyperplanes that separates regions in the configuration
space. However in the large time and hence large points limit these thicknesses will surely become irrelevant. Sure, when the 
hyperplane thicknesses are finite the system should take longer to reach a stable asymptotic state, yet for $p=1$ that it will reach it beyond the hyperplanes is a fact. So the model with step function captures, and in fact defines, the essentials of 
models of this type.

Furthermore any model of this type will have a hydrodynamical limit where one deals with a conservation law with probability density $\mathcal{P}$ and current $\vec{v}\mathcal{P}$ where $\vec{v}$ will be defined by the microscopic rules of the game. Thus
Eq.\ref{eq:voila} is generic to all such type models. Hence on discrete lattices the model is a generic vertex model which has
a map to a spin 1/2 model where spins live on the edges.

So for this type of models there always is a  connection between, linear hyperbolic equations, hyperplane (graph) arrangements and graph colourings. One is tempted to speculate that this connection to hyperbolic equations will shed further light on the 
other disciplines mentioned above.

The most obvious generalization that comes to mind is to study the dynamics of 3-agent games to graphs. As mentioned this model
has been extensively studies only on the complete graph. A study along this line is currently under way.

Also, since it deals with constrained colourings, our model also has connections to constrained anti-ferromagnetic Potts models. The constraints can be seen as magnetic fields that affect only a given colour. There are already some work in the litterature
along these lines.

On the two dimensional torus we have said that our model is actually related to a generic 16 vertex model. However as we have
seen, the  symmetry $n_{S}=n_{L}$, $n_{M^{+}}=n_{M^{-}}$ along with $2n_{L}+2n_{M^{+}}+n_{M^{0}}=1$ is imposed on the system and one actually has a two parameter vertex model. Since all sixteen possible orientations of the vertices are present we could not
find a solution to the Yang-Baxter equation. But, this type of vertex model might be of interest in of itself. We reserve this
for further study.

As for the applications of the 2-agent model on the graphs to various phenomena one can come up with many ideas. We have already
advocated a possible application to the water levels of connected water systems. Another application may be to apply it to the
economical relations of a group of individuals with a given connectivity pattern; not all individuals have relations with
all other individuals espescially if there are hostilities and this may define edges that will never occur in  the graph the
individuals define. Once such a pattern is defined the rest will follow from the model.

The author thanks M. Mungan and F. \"{O}zt\"{u}rk for various ideas and feedback they provided regarding this work.

\appendix*\label{app:bir}

\section{The chemical potential approach for the ring graph}

In this appendix we shall develop an approximation for the statistics of our model on the ring graph. We shall assume that all
configurations with the same number of M-type agents are equally probable. But we shall weigh them with respect to the full
LS checkerboard configuration via a chemical potential $\mu$. For instance if a configuration contains $k$ many M-type agents we shall weigh it with $\mu^{k}$.

As well known, to obtain the statistics of the ring diagram one will need to solve the path diagram first. But since we shall ultimately be considering the limit of infinite number of vertices we do not really need to solve for the ring diagram's partition
function: Using a well known trick, whenever we need to calculate the expectation value of a given finite size subconfiguration we can place it in the middle of an infinitely long path diagram.

Now let us remember that the configurations we are interested in  are all 3-colourings where the colours are labelled as $L$, $S$ and $M$ respectively and where tiles of type $LML$ and $SMS$ are forbidden.

We therefore have the following partition function for the path graph

\be
Z_{N}\equiv \sum_{\{ \sigma_{i}\}}\prod_{i=1}^{N} {\mathcal{Q}}_{\sigma_{i-1},\sigma_{i},\sigma_{i+1}},
\ee
where the measure $\mathcal{Q}$ is given as
\be
{\mathcal{Q}}_{\sigma_{i-1},\sigma_{i},\sigma_{i+1}}=(1-\delta_{\sigma_{i},\sigma_{i+1}})(1-(\mu-1)\delta_{\sigma_{i},M})(1-\delta_{\sigma_{i-1},L}\delta_{\sigma_{i},M}\delta_{\sigma_{i+1},L})(1-\delta_{\sigma_{i-1},S}\delta_{\sigma_{i},M}\delta_{\sigma_{i+1},S}).
\ee

Summing over the first agent we get the following recursion relation

\be
Z_{N}=(\mu+1)Z_{N-1}-\mu M_{N}Z_{N},
\ee
where
\be
M_{N}Z_{N}\equiv \sum_{\{ \sigma_{i}\}}\prod_{i=1}^{N} \delta_{\sigma_{1},M}{\mathcal{Q}}_{\sigma_{i-1},\sigma_{i},\sigma_{i+1}},
\ee
and thus $M_{N}$ is the probability that the first agent on the path has colour $M$. Summing over the first agent we get the following reursion relation

\be
M_{N}Z_{N}=\mu\left(1-M_{N-1}\right)Z_{N-1},
\ee

Let us also define the following, since we shall eventually need them
\begin{subequations}
\bea
L_{N}Z_{N} &\equiv & \sum_{\{ \sigma_{i}\}}\prod_{i=1}^{N} \delta_{\sigma_{1},L}{\mathcal{Q}}_{\sigma_{i-1},\sigma_{i},\sigma_{i+1}},\\
S_{N}Z_{N} &\equiv & \sum_{\{ \sigma_{i}\}}\prod_{i=1}^{N} \delta_{\sigma_{1},S}{\mathcal{Q}}_{\sigma_{i-1},\sigma_{i},\sigma_{i+1}},\\
\eea
\end{subequations}
and thus $L_{N}$ and $S_{N}$ are the probabilities that the first agent has colour $L$ and $S$ respetively. Since these are exclusive we of course have $M_{N}+L_{N}+S_{N}=1$.

One can, summing the first agent out, show that $L_{N}$ and $S_{N}$ obey the following

\begin{subequations}
\bea
L_{N}Z_{N} &=& Z_{N-1}-L_{N-1}Z_{N-1}-\mu L_{N-1}Z_{N-2},\\
S_{N}Z_{N} &=& Z_{N-1}-S_{N-1}Z_{N-1}-\mu S_{N-1}Z_{N-2}.\\
\eea
\end{subequations}

One can by direct enumaration show that the initial conditions for $L_{N}$ and $S_{N}$ are identical and thus we must have $L_{N}=S_{N}$ which means that they are equal to $(1-M_{N})/2$. This is expected since
the constraints in the measure are symmetric under $L\leftrightarrow S$.

Now one can, by rearranging the equations involving $Z_{N}$ and $M_{N}$, show that they are equivalent to

\be\label{eq:partpath}
\boxed{Z_{N}=Z_{N-1}-\mu Z_{N-2}}
\ee

\be
\boxed{M_{N}=1-\frac{Z_{N-1}}{Z_{N}}\;\;\;\;{\rm and}\;\;\;\;L_{N}=S_{N}=\frac{1}{2}\frac{Z_{N-1}}{Z_{N}}}
\ee

To resolve Eq.\ref{eq:partpath} we need to provide two initial conditions. But these initial conditions must
be such that they cover all the relevant constraints and thus $Z_{3}$ must be in the list. With these we get the following solution

\be
Z_{N}=A_{+}\lambda_{+}^{N}+A_{-}\lambda_{-}^{N},
\ee
with

\begin{subequations}
\bea
\lambda_{\pm}&=&\frac{1}{2}\left(1\pm\sqrt{1+4\mu}\right),\\
A_{\pm} &=&1\pm\frac{1+2\mu}{\sqrt{1+4\mu}}.
\eea
\end{subequations}

Note that $\vert \lambda_{+} \vert>\vert \lambda_{-}\vert$ and hence we have

\begin{subequations}
\bea
Z_{N\to\infty}&=&A\lambda_{+}^N\\
M_{N\to\infty}&=&\frac{\lambda_{+}-1}{\lambda_{+}}\\
L_{N\to\infty}&=&\frac{1}{2\lambda_{+}}
\eea
\end{subequations}

\subsection{Mean number of M-type agents}

To get the average number of $M$-type agents we consider the probability that the agent in the middle of a path graph is of type $M$. This means we need to calculate,

\be
\bar{M}_{2N+1}Z_{2N+1}=\sum_{\{ \sigma_{i}\}}\prod_{i=-N}^{N} \delta_{\sigma_{0},M}{\mathcal{Q}}_{\sigma_{i-1},\sigma_{i},\sigma_{i+1}}
\ee

By summing over the agent $\sigma_{0}$ we shall end up with

\be
\bar{M}_{2N+1}Z_{2N+1}=2\mu L_{N}^{2}Z_{N}^{2}
\ee

In the limit $N\to\infty$ the quantity $\bar{M}_{2N+1}$  must converge to $n_{M}$ of the paper and we recover

\be
\boxed{n_{M}=\frac{1}{2}\left(1-\frac{1}{\sqrt{1+4\mu}}\right)}
\ee
as advertized before via other means.

\subsection{The distance distribution between two M-type agents}

To study the probability distribution $P(k)$ of distances between two M-type agents we again place the region of interest in the middle of a path graph of length $2N+k+2$ with $k\geq 1$. A straightforward calculation yields

\be
P(k)=\frac{2}{Z_{2N+k+2}}\mu^{2}L_{N}^{2}Z_{N}^{2},
\ee 
which in the large $N$ limit will give us

\be
\boxed{P(k)=\frac{\mu^{2} A}{2\lambda_{+}^{4}}\;\lambda_{+}^{-k}}
\ee

As expected and advertized in the paper via other means this probability distribution adds up to $n_{M}$

\be
\sum_{k=1}^{\infty}P(k)=\frac{1}{2}\left(1-\frac{1}{\sqrt{1+4\mu}}\right)=n_{M}.
\ee

Furthermore again as advertized in the paper  via other means we see that
$\lambda_{+}=e^{\alpha}$ with $\alpha=\ln(2n_{L}/(4n_{L}-1))$.

\subsection{The pair correlation function of asymptotic speeds}

The quantity of interest is 

\be
C_{k+1}\equiv\langle z_{i}z_{i+k+1}\rangle=4\langle L_{i}L_{i+k+1}\rangle+2\langle M_{i}L_{i+k+1}\rangle+2\langle L_{i}M_{i+k+1}\rangle+\langle M_{i}H_{i+k+1}\rangle.
\ee

But due to the flip symmetry the middle two terms are equal and thus we have

\be
\langle z_{i}z_{i+k+1}\rangle=4\langle L_{i}L_{i+k+1}\rangle+4\langle L_{i}M_{i+k+1}\rangle-\langle M_{i}M_{i+k+1}\rangle.
\ee

Furthermore since $L_{i+k+1}+M_{i+k+1}+S_{i+k+1}=1$ we can recast this as

\be
\langle z_{i}z_{i+k+1}\rangle=4 n_{L}+\langle M_{i}M_{i+k+1}\rangle-4\langle L_{i}S_{i+k+1}\rangle.
\ee

Some tedious algebra will yield the following equations

\begin{subequations}
\bea
\langle M_{i}M_{i+k+1}\rangle &=& 2\frac{\mu^{2}L_{N}^{2}Z_{N}^{2}Z_{k}}{Z_{2N+2+k}}\left[B_{LL}(k)+B_{LS}(k)\right],\\
\langle L_{i}S_{i+k+1}\rangle  &=& \frac{Z_{K}B_{LS}(k+2)}{Z_{2N+2+k}}\left(\mu^{2}L_{N-1}^{2}Z_{N-1}^{2}+L_{N}^{2}Z_{N}^{2}+2\mu L_{N}L_{N-1}Z_{N}Z_{N-1}\right).
\eea
\end{subequations}

The quantity $B_{LL}(k)$ is the probability that a path graph starts with an L-type agent and ends with an L-type agent. Similarly the quantity $B_{LS}$ is the probability that a path graph starts with an L-type agent and ends with an S-type agent. They both satisfy the following equation

\be
f_{k}=\frac{1}{2}Z_{k-2}-f_{k-1}-\mu f_{k-2},
\ee
where $f_{k}$ stands either for $B_{LS}(k)Z(k)$ or for $B_{LL}(k)Z(k)$.

However as one can easily show that their initial conditions, which one can explicitly find by direct enumeration of graphs starting with $k=2$, are not the same. Therefor they are not equal, and in fact
there is no a priori expectation for them to be equal anyways.

One can readily find the particular solution to be $f^{P}_{k}=Z_{k}/2$. Therefor the general solution can
be cast as $f_{k}=f^{P}_{k}+g_{k}$ where $g_{k}$ obeys

\be
g_{k}=-g_{k-1}-\mu g_{k-2},
\ee
which can be readily solved to give

\be
g_{k}={\tilde{A}}_{+}{\tilde{\lambda}_{+}}^{k}+{\tilde{A}}_{-}{\tilde{\lambda}_{-}}^{k},
\ee
where one has
\be
{\tilde{\lambda}}_{\pm}=-\frac{1}{2}\left(1\pm \sqrt{1-4\mu}\right).
\ee

Note that $\vert \lambda_{+}\vert > \vert \tilde{\lambda}_{\pm}\vert$. All one now needs are the initial conditions and we have $B_{LL}(2)=0$ and $B_{LL}(3)=1/Z_{3}$ along with $B_{LS}(2)=1/Z_{2}$ and $B_{LS}(3)=\mu/Z_{3}$. 

With these and taking the $N\to\infty$ limit we finally have

\be
C_{k+1}=4n_{L}+\alpha\frac{Z_{k}}{{\lambda_{+}}}^{k}\left(B_{LL}(k)+B_{LS}(k)\right)-\beta\frac{Z_{k+2}}{{\lambda_{+}}^{k+2}}B_{LS}(k+2),
\ee
with 

\begin{subequations}
\bea
\alpha &=& \frac{\mu^{2}}{2{\lambda_{+}}^{4}}A_{+},\\
\beta &=& A_{+}.
\eea
\end{subequations}

We have thus ahieved an exact expression for the pair correlation functions in the limit $N\to\infty$. One can furthermore show that as $k\to\infty$ one has $C_{k}\to 1$. This expected since $\langle z_{i}z_{i+k}\rangle$ is expected to factorize as $\langle z_{i}\rangle\langle z_{i+k}\rangle$ and  $\langle z_{i}\rangle=1$ via the constraints on the system.

\end{document}